\RequirePackage{fix-cm}

\documentclass[10pt]{article}
\usepackage[left=3cm,right=3cm,top=2.25cm,bottom=2.25cm]{geometry} 
\usepackage{amsmath,amsfonts,amssymb,mathtools}
\usepackage{graphicx}
\usepackage{color} 
\usepackage{subcaption}
\usepackage{tabularx}
\usepackage{hyperref}
\usepackage{booktabs}
\usepackage{enumitem}
\usepackage[normalem]{ulem} 
\usepackage{multirow}
\usepackage{graphicx}

\newcommand{\MAPest}{\hat{\theta}^\mathrm{MAP}}
\newcommand{\MAP}{\mathrm{MAP}}
\newcommand{\thetasample}{\theta^{(s)}}
\newcommand{\tildethetasample}{\tilde{\theta}^{(s)}}

\newcommand{\sweight}{w^{(s)}}
\newcommand{\unsweight}{\tilde{w}^{(s)}}

\newcommand{\thetaTV}{\theta^\textrm{TV}}
\newcommand{\hatthetaTV}{\thetaTV}
\newcommand{\yn}{y_{1:n}}

\newcommand{\rd}{\mathrm{d}}
\newcommand{\iid}{\mathrm{iid}}
\newcommand{\cov}{\text{cov}}
\newcommand{\Sample}{\mathbb{S}}

\makeatletter
\newcommand{\objref}[4]{\def\obj@rg{#4}%
  #1\ifx\obj@rg\empty#2\else#3\xspace\ref{#4}--\fi\ref}
\makeatother
\newcommand{\Sobjref}[1]{\objref{#1}{~}{s}}

\newcommand{\Figref}[1][]{\Sobjref{Figure}{#1}}

\newcommand{\Eqref}[1][]{\Sobjref{Eq.}{#1}}

\usepackage[numbers]{natbib}
\usepackage[english]{babel}
\usepackage{float}
\usepackage{bbm}
\usepackage{amsmath,amsfonts,amssymb,mathtools}
\usepackage{epstopdf}
\usepackage{url}
\usepackage{hyperref}
\usepackage{enumitem}
\usepackage{setspace}
\usepackage{multirow}
\graphicspath{{figures/}}
\usepackage{xr}
\begin{document}
\title{\bf Bayesian data assimilation to support informed decision-making in individualised chemotherapy
}


\author{
Corinna Maier$^{1,2}$, 
Niklas Hartung$^{1}$, Jana de Wiljes$^{1,3}$,\\[1ex] Charlotte Kloft$^{4}$, and
Wilhelm Huisinga$^{1,\ast}$
}



\date{}

\maketitle
$^1$Institute of Mathematics, University of Potsdam, Germany,\\[1ex]
$^2$Graduate Research Training Program PharMetrX: Pharmacometrics \& Computational Disease Modelling, Freie Universit\"at Berlin and University of Potsdam, Germany\\[1ex]
$^3$Departement of Mathematics and Statistics, University of Reading, Whiteknights, UK\\[1ex]
$^4$Department of Clinical Pharmacy and Biochemistry, Institute of Pharmacy, Freie Universit\"at Berlin, Germany\\\\[1ex]
$^\ast$corresponding author (huisinga@uni-potsdam.de)

\begin{abstract}
An essential component of therapeutic drug/biomarker monitoring (TDM) is to combine patient data with prior knowledge for model-based predictions of therapy outcomes. Current Bayesian forecasting tools typically rely only on the most probable model parameters (maximum a-posteriori (MAP) estimate). This MAP-based approach, however, does neither necessarily predict the most probable outcome nor does it quantify the risks of treatment inefficacy or toxicity. Bayesian data assimilation (DA) methods overcome these limitations by providing a comprehensive uncertainty quantification. 
We compare DA methods with MAP-based approaches and show how probabilistic statements about key markers related to chemotherapy-induced neutropenia can be leveraged for more informative decision support in individualised chemotherapy. 
Sequential Bayesian DA proved to be most computational efficient for handling interoccasion variability and integrating TDM data. For new digital monitoring devices enabling more frequent data collection, these features will be of critical importance to improve patient care decisions in various therapeutic areas.



\end{abstract}


\section*{INTRODUCTION}\label{sec:Intro}
%


In the presence of a narrow therapeutic window and large inter-patient variability, therapeutic drug/ biomarker monitoring (TDM) is indicated for safe and efficacious therapies. With the help of Bayesian forecasting tools, patient-specific data are combined with prior knowledge from previous clinical studies and a drug-specific model to enable model-informed precision dosing (MIPD) \cite{Keizer2018}. Typically, only the most probable individual parameter values, i.e., the maximum a-posteriori (MAP) estimates, are used to predict the individual therapy outcome without quantifying associated uncertainties \cite{Sheiner1979}. Thus, relevant risks associated with a dosing regimen selection, e.g. treatment inefficacy or unacceptable toxicity, are not determined hindering a well-founded therapeutic decision-making.

In this article, we thoroughly compare in a TDM context Bayesian data assimilation (DA) methods that allow for a comprehensive uncertainty quantification by estimating the full posterior distribution (termed full Bayesian approach) in contrast to MAP estimation and a normal approximation (NAP) to the posterior at the MAP estimate. The full Bayesian approaches comprise not only methods that process patient data collected over time in a batch (i.e., all at once), like Markov chain Monte Carlo (MCMC) and sampling importance resampling (SIR), but also particle filters (PF) that allow for efficient sequential data processing. PF are well established in areas of application in which real-time predictions based on online/monitoring data are required, as in navigation, meteorology and tracking \cite{ReichCotter2015,Shuman1958,Mihaylova2014}. 

In the context of chemotherapy-induced neutropenia---the most frequent dose-limiting side effect
for cytotoxic anticancer drugs with substantial decrease of neutrophil granulocytes and thus potentially life-threatening fever and infections \cite{Crawford2004}---we demonstrate the clear benefits of uncertainty quantification compared to purely MAP-based predictions (as, e.g., in \cite{Netterberg2017}) using the gold-standard model for neutropenia \cite{Friberg2002}. Further, we compare the full Bayesian approaches regarding quality of uncertainty quantification and computational runtime for multiple cycle chemotherapy \cite{Henrich2017}. While MCMC, SIR and PF all provide a reliable uncertainty quantification, the efficient data processing of the sequential approach will be clearly beneficial in a continuous monitoring context, where digital health care devices (e.g. wearables) allow patients to measure and report individual marker concentrations online.


\section*{METHODS}\label{sec:Methods}


First, the statistical framework of MIPD in TDM is introduced, which is used throughout the different methods described below. Then, the considered clinical application scenarios are described along with the prior knowledge from literature.

\subsection*{\bf Statistical framework}
TDM in the context of MIPD builds on prior knowledge in form of a structural, observational, covariate and statistical model. In the sequel, TDM data are considered for a single individual and therefore there is no running index for individuals. The structural and observational models are given as 
\begin{align}
\frac{\rd x}{\rd t}(t) &= f(x(t);\theta,u), \qquad x(t_0) =x_0(\theta)  \label{eq:ode} \\
h(t) &= h(x(t),\theta) \label{eq:observable} %
\end{align}
with state vector $x=x(t)$ (incl.\ drug/biomarker concentrations), individual parameter values $\theta$ (e.g., volumes, clearances) and rates of change $f(x;\theta, u)$ of all state variables for a given input $u$ (e.g., dose). Since typically only a part of the state variables is observed, the function $h$ maps $x$ to the observed quantities $h(x,\theta)$, e.g., plasma drug or neutrophil concentration, including potential state-space transformations (e.g., log-transformed output). The initial conditions $x_0$ are defined by the pre-treatment levels (e.g., baseline values). The covariate and statistical model link the patient-specific covariates `$\cov$' and observations $(t_j, y_j)_{j=1,\ldots,n}$ to the model predictions $h_j(\theta) = h(x(t_j),\theta)$, accounting for measurement errors and possibly model misspecification,
\begin{align}
Y_{j|\Theta=\theta} &\sim p\big(\,\cdot\, | \theta; h_j(\theta),\Sigma\big)\,, \qquad j=1,\dots,n\quad (\text{independent})\label{eq:ResidualModel}\\
\Theta &\sim p_\Theta\big(\,\cdot\,; \hatthetaTV(\cov),\Omega\big)\,, \label{eq:IIVModel}%
\end{align}
where $\hatthetaTV(\cov)$ denotes the typical hyper-parameter values (TV) that might depend on covariates. The dot (`$\cdot$') in a probability distribution serves as placeholder for its argument.
Often, $Y_{j|\Theta=\theta} = h_j(\theta) + \epsilon_j$ with $\epsilon_j \sim_\iid \mathcal{N}(0,\Sigma)$. Prior knowledge about the parameters is provided by population analyses of clinical studies, in which nonlinear mixed effects (NLME) approaches are used to estimate the functional relationship $\cov\mapsto\thetaTV(\cov)$, and the parameters $\Sigma$ and $\Omega$. 

The challenge in MIPD is to infer information on the individual parameter values $\theta$ of a patient based on his/her covariate values and measurements. Here, a Bayesian approach is highly beneficial: the unexplained inter-individual variability in the population model (\Eqref{eq:IIVModel}) defines the prior uncertainty about the individual parameter values. In this context, the hyperparameters, i.e., all parameters after the semicolon in \Eqref{eq:IIVModel}, are assumed to be known (fixed). As a consequence, we drop them as well as the subscripts in the notation in the sequel. As a result, the likelihood at the individual level reads $p(\cdot|\theta) = p(\cdot | \theta; h_j(\theta),\Sigma)$ and the prior $p(\cdot) = p_\Theta(\,\cdot\,; \hatthetaTV(\cov),\Omega)$. Then, assimilating measurements $\yn=(y_1,\dots,y_n)^T$ into the model based on Bayes' formula
\begin{equation} \label{eq:Bayes}
p(\theta|\yn) = \frac{ p(\yn|\theta)\cdot p(\theta)}{p(\yn)}
\end{equation}
allows to learn about individual parameter values from the data. The remaining uncertainty of parameter values is encoded in the posterior $p(\cdot |\yn)$. Note that $p(\yn|\theta) = p(y_1|\theta)\cdot\ldots\cdot p(y_n|\theta)$ due to independence in \Eqref{eq:ResidualModel}. The denominator $p(\yn)$ in \Eqref{eq:Bayes} serves as a normalisation factor, denoting the probability of the data. 
In contrast to MAP estimation, which summarises the posterior by its mode, Bayesian DA approaches rely on sample approximations of the posterior
\begin{equation}\label{eq:EmpiricalMeasure}
p(\theta|\yn) \approx \sum_{s=1}^S \sweight_n \delta_{\thetasample_n}(\theta )
\end{equation}
based on a sample
\begin{equation*}
\Sample_n := \Big\{ \big(\thetasample_n,\sweight_n\big),\quad s=1,\ldots,S \Big\}\,,
\end{equation*}
with sample parameters ${\thetasample_n}$ of the posterior \cite[section 2.5]{Sarkka2013}, weights $\sweight_n$, which sum to one, i.e. $\sum_{s=1}^S \sweight_n$=1 and point masses $\delta_{\thetasample_n}$ at ${\thetasample_n}$. If $\sweight_n=1/S$ for all $s$, the sample is called unweighted. Based on the posterior sample, we may also approximate quantities of interest in the observable space by solving \Eqref{eq:ode}+\ref{eq:observable} for all elements in $\Sample_n$. This serves also as the basis for credible intervals; applying subsequently the residual error model  \Eqref{eq:ResidualModel} is the basis for prediction intervals.

Since direct sampling from the posterior is in general not possible, alternative approaches (described below) need to be employed to generate a sample of the posterior. For a detailed description, see Section~S~3 - S~7. 
\\




\subsection*{\bf Maximum a-posteriori (MAP) estimation}

MAP estimation approximates the mode of the posterior distribution, i.e., the most probable parameter values given patient-specific measurements $\yn$:
\begin{align}\label{eq:MAP_objFct}
\hat{\theta}^\MAP_n = \underset{\theta}{\arg \max} \ p(\theta|\yn).
\end{align}
The MAP estimate $\hat{\theta}^\MAP_n$ is a one-point summary of the posterior distribution, without quantification of the associated uncertainty.


\subsection*{\bf Normal approximation (NAP)}

To overcome the one-point summary limitation of the MAP estimate, the posterior $p(\cdot|\yn)$ may be approximated locally by a normal distribution located at the MAP estimate \cite[section 4.1]{Gelman2014}
\begin{equation}\label{eq:NAP}
p(\cdot|\yn) \approx \mathcal{N}\Big( \hat{\theta}^\MAP_n,\,  \mathcal{I}^{-1}\big(\hat{\theta}^\MAP_n\big) \Big)\,,
\end{equation}
where $\mathcal{I}$ denotes the total observed Fisher information matrix \cite[section 2.5]{BoosStefanski2013} 
\begin{displaymath}
\mathcal{I}\big(\theta \big) := \mathcal{I}^\text{post}\big(\theta \big) = - \frac{d^2}{d\theta^2} \log p(\theta|\yn)
\end{displaymath}
of the posterior. The uncertainty in the parameters is then propagated to the observables by first sampling from the normal distribution in \Eqref{eq:NAP} and then solving the structural model for each sample \cite{Kummel2018}. Alternatively to this sampling-based approach, the Delta method \cite[section 5.5]{Wasserman2000} could be used, see also Section~S~4. 


\subsection*{\bf Sampling Importance Resampling (SIR)}

The SIR algorithm is a full Bayesian approach. It generates an unweighted sample $\Sample_n$ from the posterior $p(\theta|\yn)$ based on a sample from a so-called importance distribution $G$, from which samples can easily be generated \cite[section 10.4]{Gelman2014}. SIR proceeds in three steps. S-Step: iid.\ sampling from the importance distribution $G$ resulting in a sample $\tilde\Sample_n$. We considered the prior as importance distribution, assuming that the patient under consideration is sufficiently well represented by the clinical patient populations given by the prior. I-Step: Each sample point $\tilde{\theta}^{(s)}_n\in\tilde\Sample_n$ is assigned an importance weight $\unsweight_n = p\big(\yn | \tilde{\theta}_n^{(s)} \big)$ given by the likelihood (in case $G$ is the prior). R-Step: After normalisation of the weights $\sweight_n = \unsweight_n/ \sum_s \unsweight_n$, a resampling is performed: $S$ unweighted samples $\thetasample_n$ are drawn from $\tilde{\theta}^{(s)}_n$ according to weights $\sweight_n$.

Note that once a new data point $y_{n+1}$ becomes available, the SIR algorithm does not simply update the present sample points $\thetasample_n$ in $\Sample_n$, but re-performs all three steps based on the updated posterior $p\big(\cdot|y_{1:n+1} \big)$ to determine $\Sample_{n+1}$.


\subsection*{\bf Markov chain Monte Carlo (MCMC)}

A popular alternative to SIR in Bayesian inference are MCMC methods with a wide range of different algorithms, see e.g. \cite{Ballnus2017}. MCMC generates an unweighted sample $\Sample_n$ from the posterior by means of a Markov chain \cite[section 11]{Gelman2014}. MCMC comprises two steps to generate sample points $\thetasample_n$: a proposal step (generating a potential new sample point $\theta^*$) and an acceptance step (accepting or rejecting $\theta^*$ as a new sample point). The challenge in MCMC is to design application-specific proposal distributions.   

In TDM, MCMC was previously considered with the prior as fixed proposal distribution (independence sampler) for sparse patient monitoring data \cite{Wakefield1996}. We observed, however, large rejection rates with increasing number $n$ of data points, since the posterior becomes narrower, see Section~S~8. To counteract large rejection rates, we used an adaptive Metropolis-Hastings sampler with log-normally distributed proposal distribution (see Section~S~6  for details). 



\subsection*{\bf Particle filtering (PF)}

In contrast to SIR and MCMC, which process data in a batch, PF constitutes a sequential approach to DA, see \cite{Gordon1993,ReichCotter2015,Arulampalam2002} for a detailed introduction. Given a weighted sample $\Sample_{n}$ of the posterior $p(\cdot | y_{1:n})$ and a new data point $y_{n+1}$, PF generates a weighted sample $\Sample_{n+1}$ of the posterior $p(\cdot | y_{1:n+1})$ by updating  
$\Sample_{n}$ using a sequential version of Bayes formula
\begin{equation} \label{eq:seqBayes}
p(\theta|y_{1:n+1}) \propto p(y_{n+1}|\theta) \cdot p(\theta| y_{1:n}).
\end{equation}
For $n=0$, the distribution $p(\theta|y_{1:0})$ is identical to the prior $p(\theta)$ \cite[section 3]{Sarkka2013}. Note that $p(y_{n+1}|\theta) =p(y_{n+1}|y_{1:n},\theta)$ due to independence in \Eqref{eq:ResidualModel}. Analogously to the I-step in SIR, the weights $\sweight_{n}$ are updated proportional to the (local) likelihood:  $\sweight_{n+1} \propto p(y_{n+1}|\theta)\cdot \sweight_{n}$ involving, however, only the new data point $y_{n+1}$. As in SIR and MCMC, evaluation of the likelihood involves solving the structural model \eqref{eq:ode}. Since the structural model is deterministic, one may either solve \Eqref{eq:ode} with initial condition $x_0(\theta)$ for the total timespan $[t_0,t_{n+1}]$, or with initial condition $x_{n}(\theta)$ for the incremental timespan $[t_{n},t_{n+1}]$. The latter approach requires to store for each sample point $(\thetasample_n,w_n^{(s)})$ also the corresponding state $x_{n}^{(s)}$ at time $t_{n}$, since typically the structural model cannot be solved analytically. The incremental approach makes use of the Markov property that the future state is independent of the past when the present state is known.
 
The resulting triple $\big(\thetasample_n, \sweight_n,x_n^{(s)}\big)$ is called a particle. In our setting, the ensemble of particles can be interpreted as the state of a  population of virtual individuals at time $t_n$, whose ``diversity" represents the uncertainty about the state/parameters of the patient at time $t_n$, given the individual measurements $\yn$. The posterior $p(\cdot | \yn)$ obtained by $n$ sequential update steps in \Eqref{eq:seqBayes} is mathematically identical to the posterior obtained in \Eqref{eq:Bayes} by assimilating all data $y_{1:n}$ in a batch \cite[section 3.3.3]{Murphy2012}. However, the sequential update is much more efficient as it involves a reduced integration time span. 

In contrast to SIR, PF does not perform resampling by default. Only if too many samples carry an almost negligible weight and the total weight is limited to only a few samples (weight degeneracy), a resampling is performed. We used a criterion based on the effective sample size
\begin{equation*}
S_\text{eff}(t_n) := \frac{1}{ \sum_{s=1}^S \Big({\sweight_n}\Big)^2}
\end{equation*}
to decide whether to resample. Starting initially with uniform weights $\sweight_0=1/S$ with $S_\text{eff}(t_0) = S$, resampling was carried out once $S_\text{eff}<S/2$ (effective ensemble size half of the initial ensemble size). If resampling is performed, it is followed by a so-called rejuvenation step  \cite{ReichCotter2015} to prevent sample impoverishment by fixation to limited parameter values: 
\begin{equation*}
\thetasample_{n}=\tildethetasample_n+ \xi^{(s)}_{n}\,, \qquad \text{with} \ \xi^{(s)}_{n} \sim_\iid \mathcal{N}(0,\tau \cdot |\tildethetasample_n|)
\end{equation*}
with rejuvenation parameter $\tau$, where $\tildethetasample_n$ denotes the resampled parameters. These two steps, resampling and rejuvenation, ensure that the weighted sample $\Sample_n$ adequately represents areas of posterior probability.


\subsection*{\bf Biomarker data during chemotherapy for single/multiple cycle simulation studies} 
\label{sec:MultipleCycle}

Two simulation studies (see below) were performed to analyse the approaches regarding their suitability to support MIPD. 

For the single cycle study with docetaxel ($100\,\text{mg/m}^2$, 1\,h infusion), we used the NLME model in \cite{Kloft2006}. It is based on the well-known pharmacodynamic (PD) model in \cite{Friberg2002} (Figure~S~6) and describes the effect of a single dose of the anticancer drug docetaxel  based on monitoring neutrophil counts. Important model paramters are the drug effect parameter `Slope' and the pre-treatment baseline neutrophil concentration `Circ0'. For inference, neutrophil concentrations were considered on a log-scale at time points $t=0,3,\dots,21$ days post-dose. See Section~S~8  for full details. This simulation study aims to demonstrate the limitations of MAP estimation for a model frequently used in MIPD for TDM (see, e.g. \cite{Netterberg2017,Wallin2009}).
Since recursive data processing and decision-support gain in relevance for long-term monitoring, we performed a simulation study for multiple cycle therapy with paclitaxel using the NLME model in \cite{Henrich2017}. It describes the effect of the anticancer drug paclitaxel ($200\,\text{mg/m}^2$, 3\,h infusion) over 6 cycles of 3 weeks each, corresponding to treatment arm A of the CEPAC/TDM study \cite{Joerger2016}. In \cite{Henrich2017}, an aggravation of neutropenia over subsequent treatment cycles is accounted for by bone marrow exhaustion \cite{Henrich2017} (Figure~S~6). The model includes interoccasion variability (IOV) on pharmacokinetic (PK) parameters describing the variability between cycles within one patient. Therefore, the parameter values comprise the interindividual parameters ($\theta_\text{IIV}$) and a parameter for each occasion ($\theta_\text{IOV}$). As a consequence, the size of $\theta$ increases with every occasion/cycle: $\theta = (\theta_\text{IIV},\theta_\text{IOV}^1,\ldots,\theta_\text{IOV}^{n_c})$, where $n_c$ denotes the number of cycles, see Section~S~9  for details. Neutrophil counts were assumed to be monitored every third day. 

We were interested in a setting where data become available sequentially (one-by-one). To this end, neutrophil count data were simulated for a virtual patient using \Eqref{eq:ResidualModel}-\ref{eq:IIVModel} and the corresponding model. Then the individual parameter values were inferred based on the simulated neutrophil count data available up to a certain time point, using the same model.  For the statistical analysis, this procedure was repeated for $N=100$ virtual patients (with covariate characteristics mirroring the real study population underlying the NLME model).


\begin{figure}
\centering
\includegraphics[width =1\linewidth]{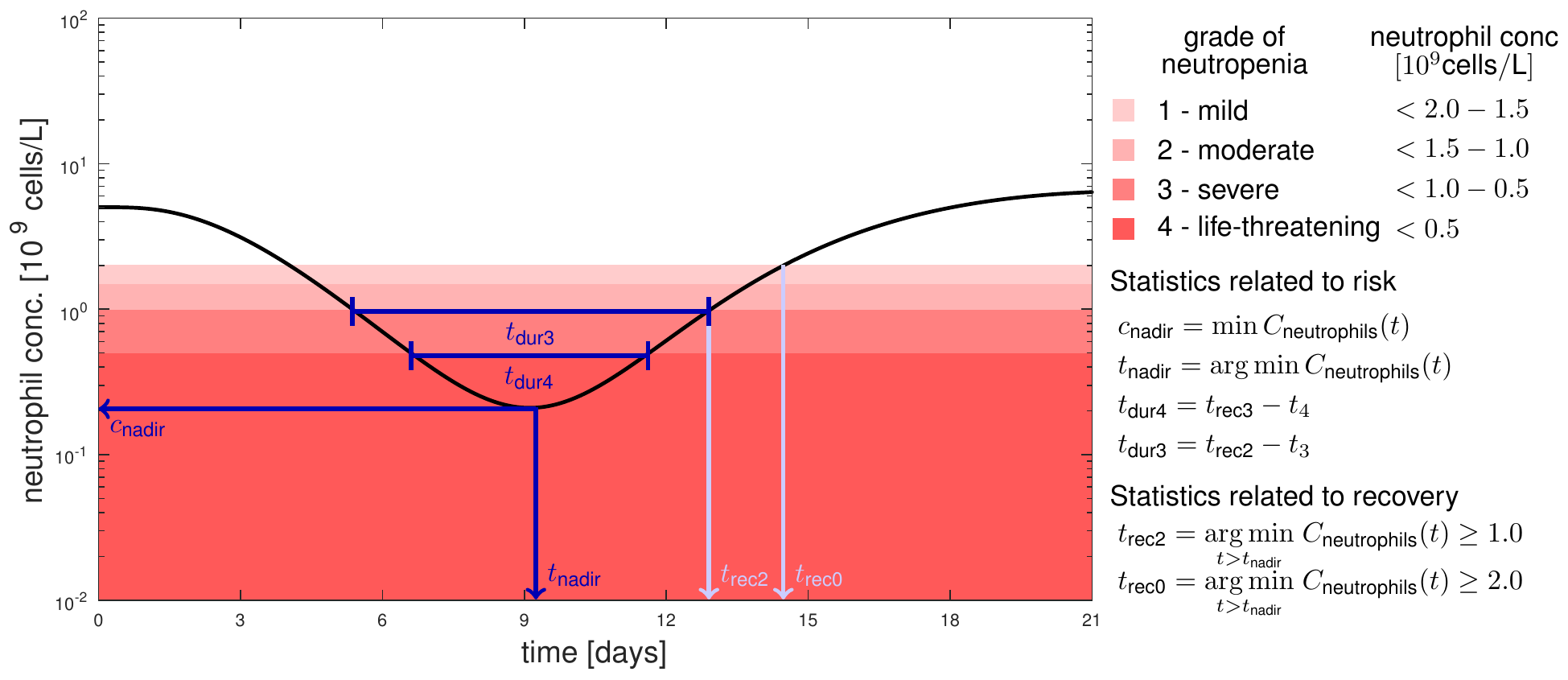}
\caption{\textbf{Key characteristics for decision-making in cytotoxic chemotherapy} related to risk (dark blue) and recovery (light blue) of neutropenia. Neutropenia grades are defined according to the Common Terminology Criteria for Adverse Events (CTCAE) \cite{CTCAE}. Note that the shades of red are related to the increasing toxicity, however, grade 0 (white) over the whole cycle is associated with ineffective treatment. As key statistics for decision support we consider the lowest neutrophil concentration ($c_\text{nadir}$), the time at which the nadir is reached ($t_\text{nadir}$), the duration of neutropenia grade 3 and grade 4 ($t_\text{dur3}$ and $t_\text{dur4}$,respectively), as well as the times until recovery to neutropenia grade 2 and 0 ( $t_{\text{rec}2}$ and  $t_{\text{rec}0}$, respectively).}
\label{fig:statistics}
\end{figure}

\subsection*{\bf Key characteristics for decision-support in cytotoxic chemotherapy}\label{sec:Statistics}

We investigated different characteristics $T(\cdot)$ of the neutropenia time-course related to risk and recovery \cite{Netterberg2017}.
Depending on the nadir, i.e., minimal neutrophil concentration, different grades of neutropenia are distinguished, see \Figref{fig:statistics}. Neutropenia grade 4 is dose-limiting as this severe reduction in neutrophils exposes patients to life-threatening infections. On the contrary, neutropenia grade 0 is also undesired as it is associated with a worse overall treatment outcome \cite{DiMaio2006a}. The time $t_\text{nadir}$, at which the nadir is reached, is important for time management of intervention. We considered the patient out of risk at time $t_{\text{rec}2}$, when neutropenia grade 2 is reached post nadir. For the initiation of the next treatment cycle, the recovery time $t_{\text{rec}0}$ to grade 0 is important. In addition, risk is also related to the duration $t_\text{dur3}$ and $t_\text{dur4}$ of an individual being in grade 3 and 4 neutropenia, respectively.


\begin{figure}
\centering
\includegraphics[width =\linewidth]{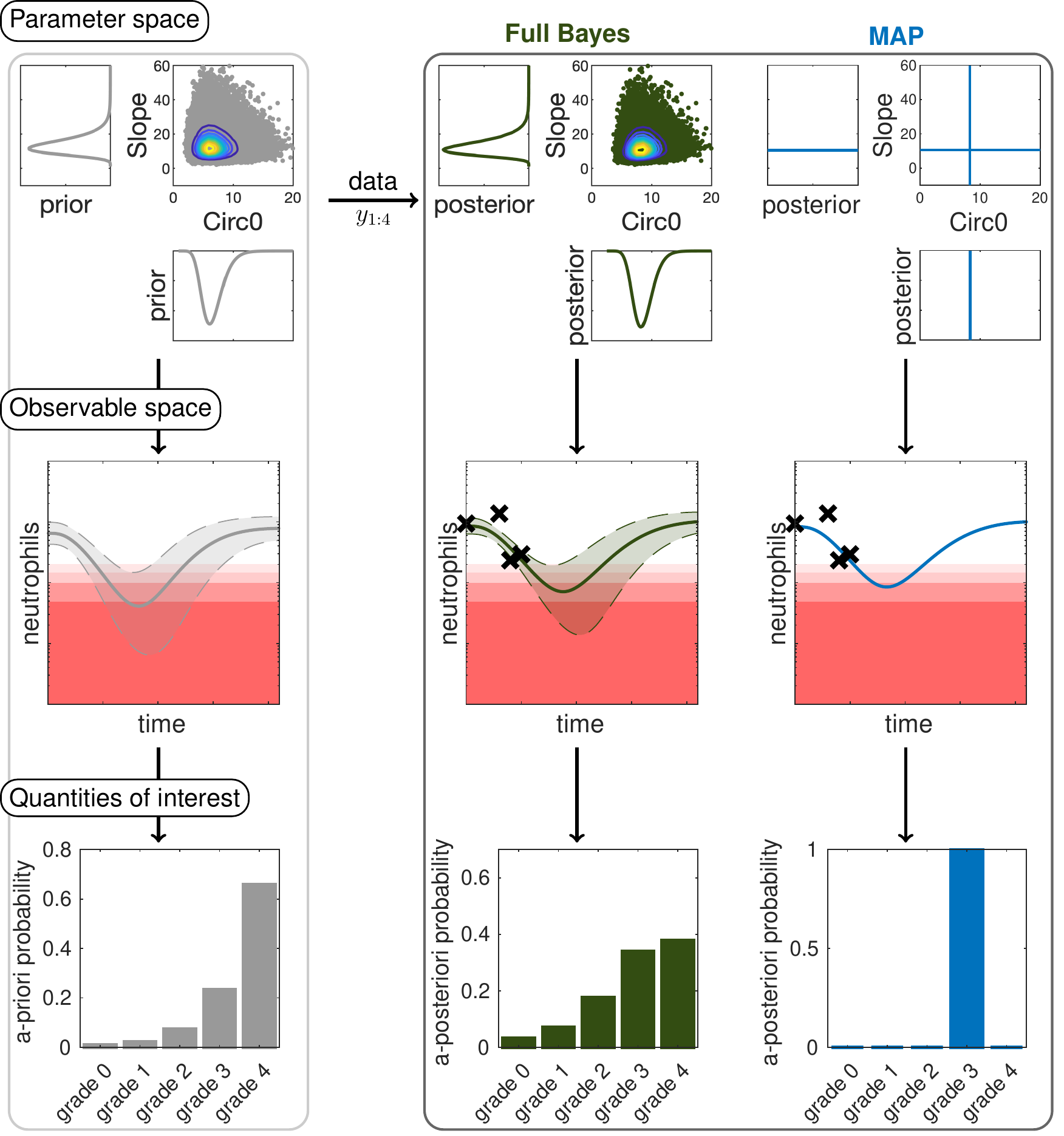}
\caption{\textbf{Overview of the workflow in MIPD comparing full Bayesian inference to MAP-based prediction}. In full Bayesian inference uncertainties in the parameter values are propagated to uncertainties in the observable space and quantities of interest. The posterior is displayed for the parameters 'Slope' (drug effect parameter) and 'Circ0' (pre-treatment neutrophil concentration). For the prior and Full Bayes approach samples (dots) from the distributions are shown with contour levels. In the observable space the point estimates (solid lines) are displayed with the central $90\%$ confidence (CI) or credible intervals (CrI) (dashed lines and shaded area) along with the TDM data (crosses). The a-priori/a-posteriori probabilities are calculated for the neutropenia grades (grade 0-4). }
\label{fig:WorkflowFigure}
\end{figure}

\subsection*{\bf Workflow in Bayesian forecasting}\label{sec:Workflow}

In full Bayesian forecasting, uncertainty is quantified on the parameter level and subsequently propagated to the observable level, possibly summarised for some key quantities of interest, see \Figref{fig:WorkflowFigure}. Prior to observing patient-specific data, the parameter uncertainty is characterised by the prior (cmp.\ \Eqref{eq:IIVModel}). It allows to make a-priori predictions of the neutropenia time course and its uncertainty in form of a $(1-\alpha)$-confidence interval. Also, a-priori predictions for quantities of interest can be derived, e.g. the neutropenia grade (\Figref{fig:WorkflowFigure}, left column). 
Once patient-specific data are assimilated into the Bayesian model, the remaining uncertainty on the parameter values is characterised by the posterior, allowing to update also the uncertainty in the observable space (credible intervals CrI) and the quantities of interest (\Figref{fig:WorkflowFigure}, middle column). 

Forward uncertainty propagation corresponds to transforming a probability distribution (prior or posterior) under a (possibly nonlinear) mapping $T(\cdot)$, resulting in a transformed quantity $\psi = T(\theta)$. For illustration, we assume the one-dimensional case with strictly increasing $T$ and $\theta=T^{-1}(\psi)$. Then the posterior in terms of $\psi$ is given by \cite[section 1]{Bishop2006}
\begin{equation}\label{eq:Transformation}
p_\Psi(\psi|y_{1:n}) =  p_\Theta(\theta|y_{1:n}) \cdot  \frac{d T^{-1}(\psi)}{d\psi}\,,
\end{equation}
which is approximated in sampling-based approaches (cf. \Eqref{eq:EmpiricalMeasure}) by 
\begin{equation*}
\hat{p}_\Psi(\psi|\yn) = \sum_{s=1}^S w_n^{(s)} \delta_{\psi^{(s)}}(\psi)\,,
\end{equation*}
with $\psi^{(s)} = T(\theta^{(s)})$. This allows the computation of any desired summary statistic, e.g. posterior expectation or quantiles. MAP estimation, in contrast, characterises the posterior by a single value and allows only to make a single MAP-based prediction by mapping the MAP estimate $\hat{\theta}^\MAP$ to the quantity of interest $T(\hat{\theta}^\MAP)$, lacking crucial information on its uncertainty (\Figref{fig:WorkflowFigure}, right column). Importantly, for nonlinear $T$ this does \textit{not} result in the most probable outcome, due to the Jacobian factor $\frac{d\theta}{d\psi} =  \frac{d T^{-1}(\psi)}{d\psi}$ in \Eqref{eq:Transformation} \cite{Jermyn2005,Lavielle2014}: The most probable outcome is defined as the outcome with maximum posterior probability
\begin{equation}
\hat{\psi}^\text{MAP}_n = \underset{\psi}{\arg \max} \ p_\Psi(\psi|\yn)\,,
\end{equation}
which satisfies (assuming for illustration that $T$ is strictly increasing)
\begin{align}\label{eq:PsiMAP}
0 = \frac{d}{d \psi} \ p_\Psi(\psi|\yn) &\overset{\Eqref{eq:Transformation}}= \frac{d}{d \psi} \left[ p_\Theta(T^{-1}(\psi)|y_{1:n}) \cdot  \frac{d T^{-1}(\psi)}{d\psi} \right] \nonumber \\
&\overset{\hphantom{\Eqref{eq:Transformation}}}= \frac{d}{d \theta}  p_\Theta(T^{-1}(\psi)|y_{1:n}) \cdot \left( \frac{d T^{-1}(\psi)}{d\psi} \right)^2 +  p_\Theta(T^{-1}(\psi)|y_{1:n}) \cdot  \frac{d^2 T^{-1}(\psi)}{d\psi^2}\,.
\end{align}
For the transformed MAP estimate $\psi = T(\hat{\theta}^\MAP)$, the first term in \Eqref{eq:PsiMAP} is zero, since its first factor vanishes by definition. The second term, however, is non-zero, since both its factors are non-zero for nonlinear $T$. Therefore, the transformed MAP estimate does not satisfy the condition for the mode of the transformed posterior probability and hence, $T(\hat{\theta}^\MAP) \neq \hat{\psi}^\MAP$.







\subsection*{\bf Method comparison }

For all sampling-based methods (NAP, SIR, MCMC, PF) we used a sample of size $S=10^3$.  Since the posterior is analytically intractable, an extensive sample of size $S=10^6$ was used as reference (generated by SIR and cross-checked with MCMC, see Figure~S~7, since these approaches are exact in the limit $S \rightarrow \infty$). 
As a statistical measure for the quality of uncertainty quantification we considered the Hellinger distance 
\begin{equation}\label{eq:Hellinger}
H(\hat{P},P^\text{ref}) := \frac{1}{\sqrt{2}} \sqrt{\sum_{i=1}^b \Big(\sqrt{\widehat{p}_i\vphantom{^\text{ref} }} - \sqrt{p_i^\text{ref} } \Big)^2} \,,
\end{equation}
which measures the difference between the discrete sampling-based a-posteriori probability distribution $\hat{P}=(\widehat{p}_1,\dots,\widehat{p}_b)$ and the reference solution $P^\text{ref}=(p^\text{ref}_1,\dots,p^\text{ref}_b)$ generated with SIR $S=10^6$ for $b$ fixed bins.


\section*{\bf RESULTS} \label{sec:Results}



First we show the limitations of MAP estimation for MIPD and how full Bayesian approaches can overcome these limitations (using SIR with $S=10^6$ for comparison). Next we compare different full Bayesian approaches with reduced sample sizes regarding accuracy and computational efficiency. 


\subsection*{\bf Unfavourable properties of MAP-based predictions} 

\begin{figure}
\centering
\includegraphics[width =0.5\linewidth]{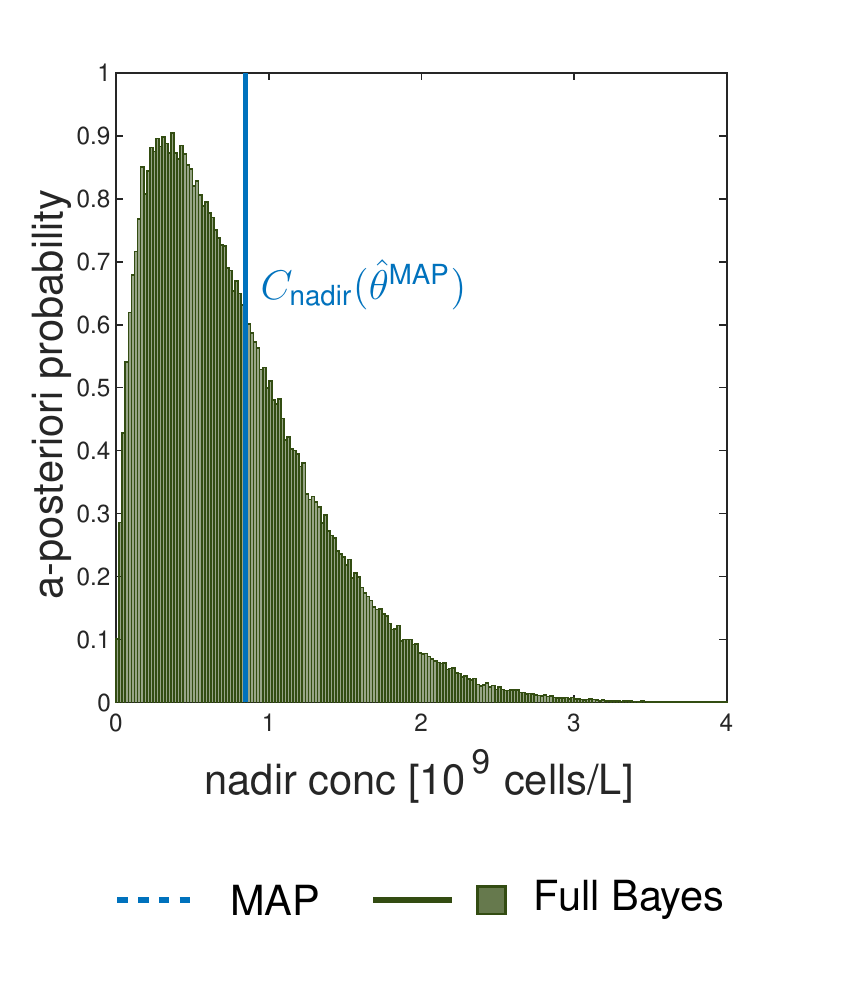}
\caption{\textbf{MAP-based predicted nadir concentration is not the most (a-posteriori) probable nadir concentration}. We considered the single cycle study docetaxel with four observed data points $y_{1:4}$ and forecasted the nadir concentration based on the posterior $p(\theta|y_{1:4})$. The mode is not preserved under nonlinear transformation (see text). Therefore, $C_\text{nadir}(\hat{\theta}^\text{MAP})$ with $C_\text{nadir}(\cdot)$ denoting some observable $T(\cdot)$ does not equal the mode of the a-posteriori probability $p_{T(\Theta)}(\cdot|\yn)$ of the nadir concentration.}
\label{fig:MAPvsFullBayes}
\end{figure}


The first example of decision support in individualised chemotherapy employs the most frequently used model of neutropenia \cite{Friberg2002}. The MAP estimate $\MAPest_n$ is derived from the parameter posterior $p(\cdot | \yn)$ given experimental data $\yn=(y_1,\dots,y_n)^T$, see \Eqref{eq:MAP_objFct}. In the context of TDM, it is used to predict the future time course $x(t;\MAPest)$ of the patient and thereon based observables. In mathematical terms, $\MAPest$ is mapped to some quantity of interest $T(\MAPest)$, e.g., the nadir concentration. As pharmacometric models are generally nonlinear, this does, however, not result in the most probable outcome (see also paragraph preceding \Eqref{eq:PsiMAP} in the Methods). This is due to the fact that first determining the MAP estimate and then applying a nonlinear mapping is in general different from first applying the mapping to the full parameter posterior and then determining its MAP estimate: $T\big(\widehat{\theta}^\MAP\big)\neq \widehat{T(\theta)}^\MAP$, see \Figref{fig:MAPvsFullBayes} for an illustration with $T(\theta)=c_\text{nadir}(\theta)$ and Figure~S~1 for more details.


Thus, MAP-based estimation lacks both, a measure of uncertainty and the feature to predict the most probable observation/quantity of interest. In addition, relevant outcomes such as the risk of grade 4 neutropenia can not be evaluated from the point estimate alone. MAP-based estimation, therefore, provides a biased basis for clinical decision-making. In contrast, full Bayesian inference provides access to the full posterior distribution of the parameters and correctly transforms uncertainties forward to the observables and quantities of interest, allowing to compute any desired summary statistic and relevant risks \cite[section 5.2]{Murphy2012}.


\subsection*{\bf Uncertainty quantifications for more comprehensive, differentiated understanding and thus better informed decision-making}

\begin{figure}
\centering
\includegraphics[width = \linewidth]{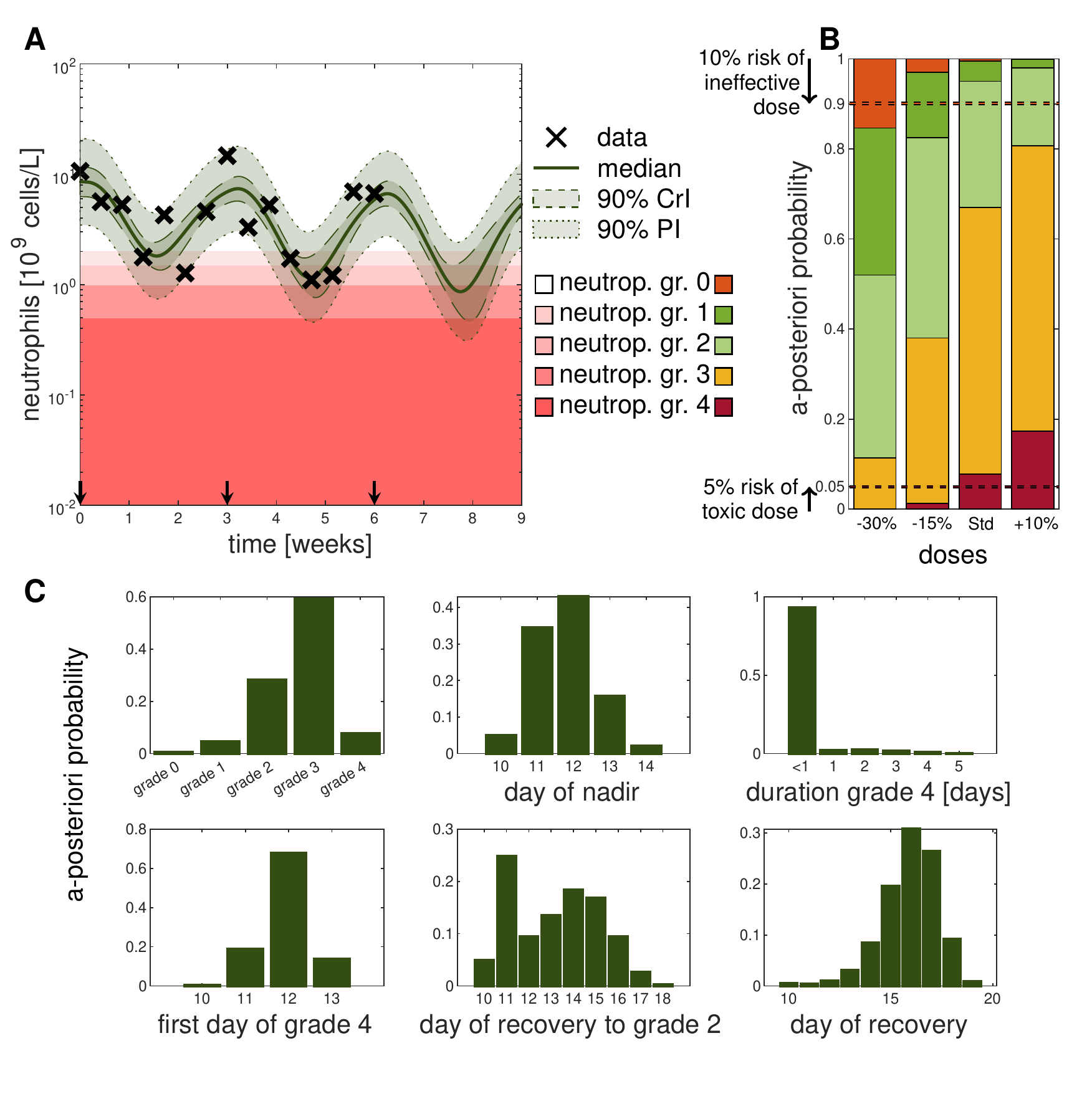}
\caption{\textbf{Uncertainty quantification by full Bayesian methods gives important information for therapy dosing selection}. The scenario described in multiple cycle study paclitaxel is used and the results are shown for the reference solution with SIR using $S=10^6$ samples. (A) Forecasting the third cycle for different doses based on the patient's covariates and measurements of the first two cycles. (B) Full Bayesian inference allows for probabilistic statements of the different grades.  Colour coding of neutropenia grades shows trade-off between efficacy and toxicity. No toxicity (grade 0) is associated with poorer treatment outcome (orange) but severe neutropenia (grade 3 and 4) is also not desired (yellow and red). (C) A-posteriori probabilities of quantities of interest for the third cycle based on the posterior at the end of second cycle (week 6) for the standard dose. Statistics such as day of grade 4 were computed given that grade 4 is reached. Note for all displayed forecasts the reference method (SIR with $S=10^6$) was used.}
\label{fig:DoseScenarios}
\end{figure}


The first scenario served to demonstrate the limitations of MAP-based estimations for the gold-standard model \cite{Friberg2002}, however, the model does not account for the observed cumulative neutropenia over multiple cycles. Therefore, we considered for dose adaptations a model accounting for bone marrow exhaustion over multiple cycles \cite{Henrich2017}, see paragraph about the multiple cycle study paclitaxel. We exemplarily considered the dose selection for the third treatment cycle based on prior information and patient-specific measurements during the first two cycles. The patient-specific data together with the full Bayesian model fit and prediction are shown in \Figref{fig:DoseScenarios}~A. The credible intervals (dashed) and prediction intervals (dotted) show the uncertainty about the `state of the patient', without and with measurement errors, respectively. 

For optimising the dose of the third cycle, different dosing scenarios were considered: the standard dose and a $-15\%$, $-30\%$ and $+10\%$ adapted dose. \Figref{fig:DoseScenarios}~B shows the probability of the predicted grades of the third cycle for each dose. To find an effective and safe dose, the risk of being ineffective (neutropenia grade 0) should be minimised jointly with the risk of being unsafe (neutropenia grade 4). For illustration in \Figref{fig:DoseScenarios}~B, the dashed horizontal lines indicate a 10\% and 5\% level of being ineffective and unsafe, respectively. The standard dose and the increased dose have a risk of toxicity larger than 5\% (lower horizontal line). A decrease in dose also leads to an increased risk of an ineffective dose (upper horizontal line). The $15\%$ reduced dose is with $96\%$ probability safe and efficacious (grade 1-3), with $3\%$ probability ineffective (grade 0) and with $1\%$ probability unacceptably toxic (grade 4).  If grade 3 is also to be avoided, the 30\% reduction would be preferable, as it is with $74\%$ probability safe and efficacious (grade 1-2), with $15\%$ probability ineffective (grade 0) and with $11\%$ probability toxic (grade 3-4). Thus, the choice of an optimal dose might depend on how priority is given to the risk of inefficacy and toxicity.
As both risks are described by the tails of the posterior distribution, a point estimate is not able to adequately capture them. The MAP-based predicted grades were: grade 2 (standard dose and $+10\%$ dose), grade 1 ($-15\%$ dose); and grade 0  ($-30\%$ dose), which do not only make it difficult to distinguish between some doses, but also do not reflect the true most probable grades.

Posterior-based predictions of important statistics related to the neutropenia time-course can help to answer questions like ``How probable is it that the patient will suffer from grade 4 neutropenia?" or ``How probable is it that the patient will recover in time for the next scheduled dose so that the therapy can be continued as planned?". To answer such questions, \Figref{fig:DoseScenarios}~C shows important predicted quantities of interest, illustrated for the standard dose in cycle 3. We inferred that the risk of grade 4 neutropenia is 8\%, and if the patient were to reach grade 4, it would be most probable ($68\%$) on day 12. The probability that the patient's duration in grade 4 is  a day or longer is very small ($<7\%$). As the probability to not have been recovered until day 21 is negligible, the administration can remain scheduled on day 21 for cycle 4. Therefore, uncertainty quantification improves the decision-making process by quantifying the a-posteriori probabilities of relevant risks and quantities of interests. Repeating the above analysis for different doses therefore allows for an improved distinction between dose adjustments.


\subsection*{\bf Approximation accuracies comparable across different full Bayesian approaches} 

\begin{figure}
\centering
\includegraphics[width = 1\linewidth]{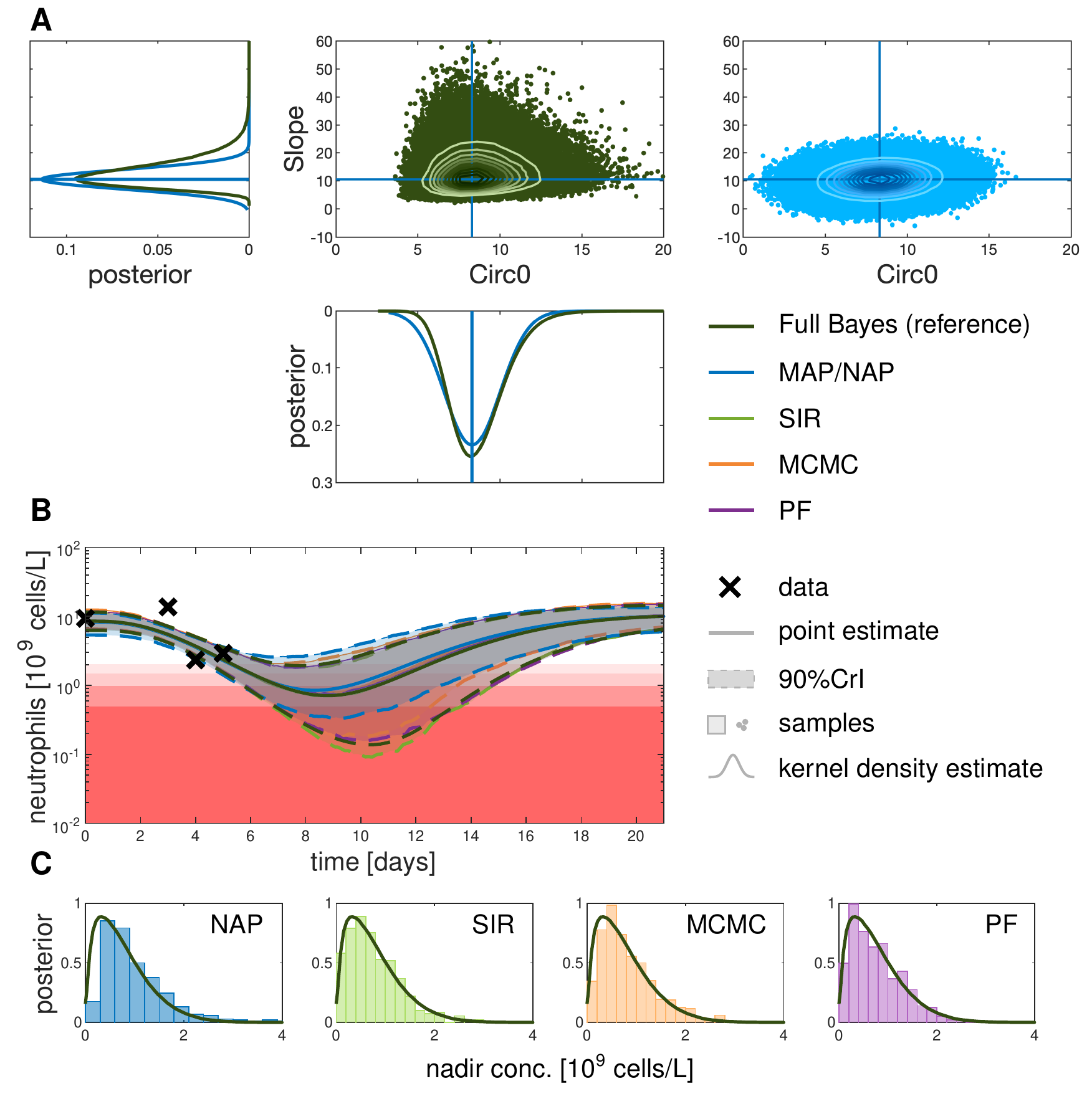}
\caption{\textbf{Comparison of uncertainty quantification at the level of parameters, observables and quantities of interest.} Exemplary comparison of the different methods for one patient after having observed four data points up to day 5. (A) The posterior is shown for parameters `Slope' and `Circ0' showing the kernel density estimates of the sampling distribution univariately and as scatter plots for the bivariate sampling distributions with contour plots for the full Bayesian approach (reference) and the normal approximation located at the MAP estimate. (B) On the level of the observable (neutrophil concentration) the point estimates (median or MAP) are displayed along with the 90\% credible intervals (CrI). For illustration purposes the prediction intervals are not shown here. (C) The forecasted a-posteriori probability of the nadir concentration is shown for the different approximations (histograms) in comparison with the reference (kernel density estimate, black solid line). }
\label{fig:UQ_Posterior}
\end{figure}

\begin{figure}
\centering
\includegraphics[width = 1\linewidth]{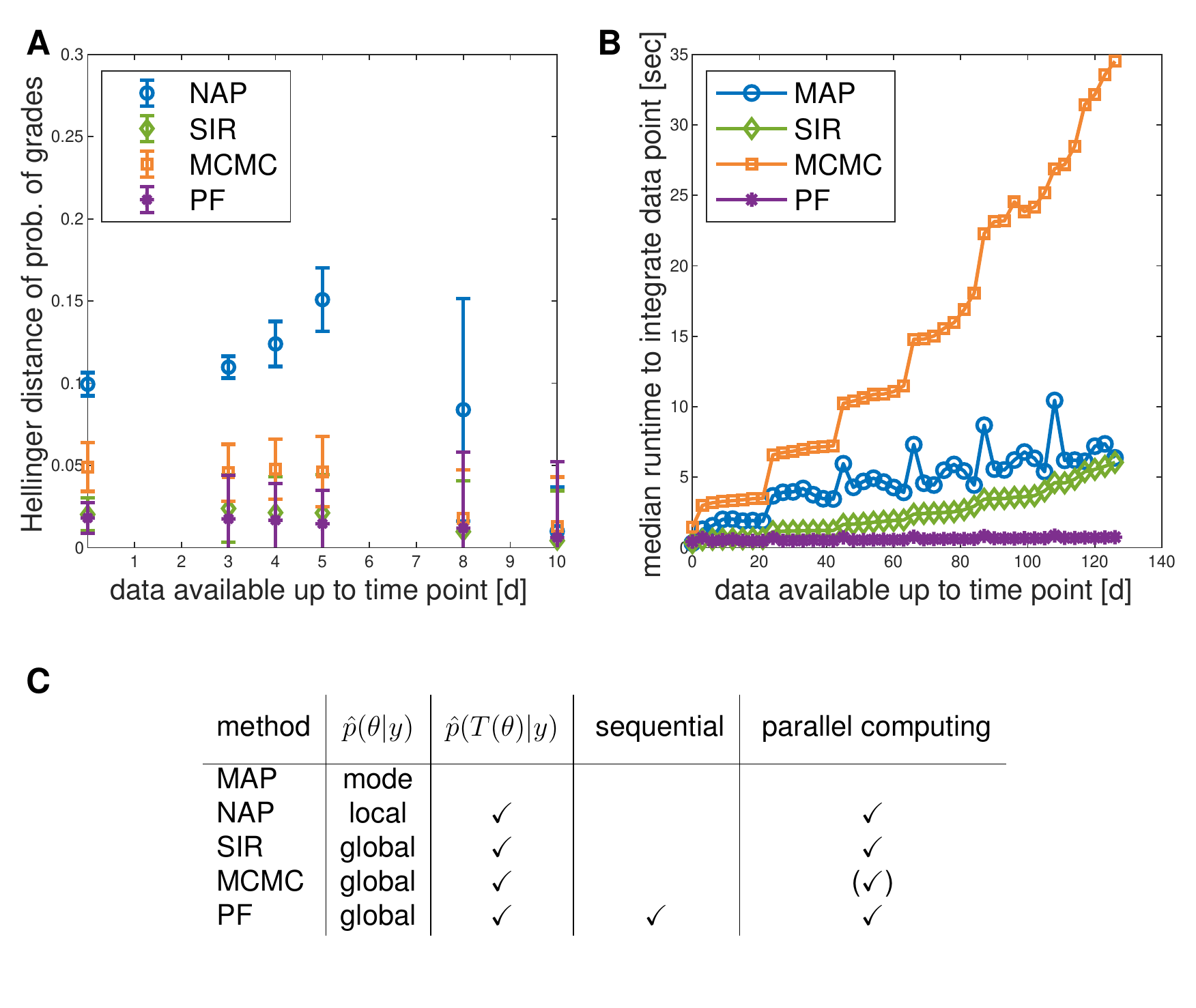}
\caption{\textbf{Comparison of methods regarding important aspects for model-informed precision dosing (MIPD)}. (A) Approximation error (measured as Hellinger distance) of the probability of neutropenia grades. (``Single cycle study docetaxel'' ). (B) Qualitative runtime comparison to sample from the parameter posterior. Median of $N=100$ repeated analyses (``Multiple cycle study paclitaxel'' ). (C) Comparison of method properties. For MCMC several chains could be run in parallel, however, in this study only one chain was considered.}
\label{fig:MethodComparisonFigure}
\end{figure}

We next compared different established methods for uncertainty quantification with regard to their approximation accuracy. To this end, posterior inference was investigated for a patient at day 5 of the first cycle (\Figref{fig:UQ_Posterior}). Whereas the marginal posterior distribution for the parameter `Circ0' (pre-treatment neutrophil concentration) is close to a normal distribution, the marginal posterior for the drug effect parameter (`Slope')  is closer to a log-normal distribution. Accordingly, the normal approximation (NAP) is rather reasonable for `Circ0', but is questionable for the `Slope' parameter. In addition, sampling from the normal distribution can lead to unrealistic (negative) parameter values (\Figref{fig:UQ_Posterior}~A). The credible intervals based on NAP underestimated the patient's risk to reach grade 4 neutropenia (\Figref{fig:UQ_Posterior}~B) as can also be seen in the posterior probability of the nadir concentration (\Figref{fig:UQ_Posterior}~C). 
Considering a Student's t distribution instead of the normal approximation, as in \cite{Kummel2018}, did not lead to an adequate improvement (Figure~S~3). Consequently, the NAP approach can result in over-optimistic, over-pessimistic and unrealistic predictions. In contrast, the full Bayesian methods (SIR, MCMC and PF) adequately represent the tails and respect the positivity constraint of parameter values. The resulting credible intervals are comparable to the reference credible intervals. 
For illustration, \Figref{fig:MethodComparisonFigure}~A shows the approximation error for the predicted probability of neutropenia grades, measured in the Hellinger distance (see \Eqref{eq:Hellinger}). Overall, SIR and PF showed the best approximation, while NAP resulted in the largest errors. 


\subsection*{\bf Sequential DA processes patient data most efficiently}

The need for real-time inference algorithms is increasing with the possibilities to more frequently collect patient-specific data (online collection) during treatment. Sequential DA methods provide an efficient framework for real-time data processing. At any time, all information (incl.\ associated uncertainty) is present in a collection of particles that can be interpreted as representing the current state and associated uncertainty of a patient via a virtual population. With a new datum, this information is updated. Approaches that rely on batch data analysis, i.e. MAP, SIR, MCMC, need to redo the inference from scratch. This has impact on the computational effort as the number of data points increases. \Figref{fig:MethodComparisonFigure}~B shows a comparison of the computational cost to assimilate an additional data point.  All approaches show some kind of  increase in effort every 21 days---due to the IOV on some parameters. Clearly, PF shows lowest and almost constant costs, while for batch mode approaches computational costs increase over time due to an increasing number of parameters (one additional parameter for every cycle due to the IOV, see paragraph about the multiple cycle study with paclitaxel) and an increasing integration time span to determine the likelihood. This could become computationally expensive in view of long term treatments and higher time resolution of data points provided by new digital health care devices. \Figref{fig:MethodComparisonFigure}~C summarises the features of the different inference approaches. Note that all sampling-based approaches can be accelerated by parallel computing. In summary, it was found that sequential DA processes patient monitoring most efficiently and facilitated the handling of IOV because only the IOV parameter of the current occasion needs to be considered.

 

\section*{Discussion}\label{sec:Discussion}



In the context of chemotherapy-induced neutropenia, we illustrated the severe drawbacks of MAP-based approaches for forecasting and thereon based decision making. A prediction based on the MAP estimate does neither correspond to the most probable outcome, nor does it allow to quantify relevant risks as the uncertainties are not quantified. Both are highly undesirable characteristics and make MAP-based inference difficult to interpret in a TDM setting. A normal approximation of the posterior at the MAP estimate is no alternative, as it retains the same point estimate and proved to be unsuitable in case of skewed parameter distributions. We demonstrated that full Bayesian approaches, like SIR, MCMC or PF provide accurate approximations to the posterior distribution, enabling comprehensive uncertainty quantification of the quantities of interest (e.g., nadir concentration). Amongst the three considered approaches, PF is a sequential approach which is beneficial in a more continuous monitoring context.

Uncertainty quantification in TDM is scarce. In \cite{Chaouch2012} the SIR algorithm was previously used in the TDM setting to construct credible intervals using a Student's t distribution located at the MAP estimate as importance function. A sequential approach in the context of MAP estimation is discussed in \cite{LeJost2018} with a moving estimation horizon (window of data points that are considered). A sequential DA approach has been investigated previously for glucose forecasting \cite{Albers2017,Albers2018}, yet not in combination with a NLME Modelling framework and without decision support statistics. A systematic comparison of approaches, as presented herein is lacking. 

In this study, particle filtering is applied in TDM within a NLME Modelling framework to represent the current patient status via an uncertainty ensemble. A challenge in the application of PF is the potential for weight degeneracy, i.e., a gradual separation into a few large and many very small weights. A rejuvenation approach (as applied in this study) resolves this problem, but requires to specify an additional parameter (magnitude of the rejuvenation). A too large value might result in an artificially increased uncertainty, while a too small value might hinder exploration of the parameter space. In the present application context, however, IOV counteracts in addition to the rejuvenation step weight degeneracy.

Sequential data processing is not only computationally efficient and convenient for IOV handling, but has the additional advantage that already assimilated experimental data need not be stored to assimilate future data points. Sampling approaches allow a simple extension for hierarchical models to include the uncertainties in the population parameters for an even more holistic uncertainty quantification. This would enable a continuous learning process between clinical trials from drug development (e.g. Phase III) and continue during the acquisition of real-world data after market authorisation, in quantifying the diverse population of patients that have taken a given drug. For a future patient, this `historic' diversity would transform into well-quantified uncertainty in a TDM setting. The absence of need to store `historic' experimental data can also be helpful for the exchange of information between clinics, health insurances  and pharmaceutical companies. The current knowledge, present in form of a sample of particles, can easily be exchanged without the need to exchange the experimental data. The `historic' data are implicitly present in the particles. 

In view of new treatments and new mobile health care devices (e.g. wearables) gathering data from various sources, clinicians have to deal with new challenges and an increasing complexity of treatment decision-making which demands for comprehensive approaches that integrate data efficiently and provide informative and reliable decision-support. We illustrated that comprehensive uncertainty quantification can result in a more informative, reliable and differentiated decision-support, which is not only limited to individualised chemotherapy but has the potential to improve patient care in various therapeutic areas in which TDM is indicated, such as oncology, infectious diseases, inflammatroy diseases, psychiatry, and transplantation patients.

\section*{Acknowledgements}

C.M. kindly acknowledges financial support from the Graduate Research Training Program PharMetrX: Pharmacometrics \& Computational Disease Modelling, Berlin/Potsdam, Germany, and from Deutsche Forschungsgemeinschaft (DFG) through grant 
CRC 1294 \lq\lq Data Assimilation\rq\rq (associated project). Fruitful discussions with Andrea Henrich (Idorsia Pharmaceuticals Ltd, Allschwil), Sven Mensing (AbbVie, Ludwigshafen) and Sebastian Reich (University of Potsdam, University of Reading) are kindly acknowledged.

\section*{Author Contributions}
C.M., N.H., J.dW., C.K., W.H. designed research, C.M. mainly performed the research, C.M., N.H., C.K., W.H. analysed data and wrote the manuscript.
\newpage
\bibliographystyle{refstyle}
\bibliography{Maier_etal_2019_literature} 



\end{document}


\maketitle
$^1$Institute of Mathematics, University of Potsdam, Germany,\\[1ex]
$^2$Graduate Research Training Program PharMetrX: Pharmacometrics \& Computational Disease Modelling, Freie Universit\"at Berlin and University of Potsdam, Germany\\[1ex]
$^3$Departement of Mathematics and Statistics, University of Reading, Whiteknights, UK\\[1ex]
$^4$Department of Clinical Pharmacy and Biochemistry, Institute of Pharmacy, Freie Universit\"at Berlin, Germany\\\\[1ex]
$^\ast$corresponding author (huisinga@uni-potsdam.de)

\tableofcontents
\newpage
\section{Introduction}
This document should help to reproduce the simulation studies (along with the supplementary MATLAB code), provide some more explanations and details and should be self-contained as many parameters and models were taken from different publications.
\section{Lognormal distribution}
Since pharmacological parameters are typically positive, e.g., volume of distribution, baseline concentrations or rate constants, often a lognormal distribution is appropriate for modelling the inter-individual variability between patients, i.e., $\theta_i = \theta^\text{TV}\cdot e^{\eta_i}$ with $\eta_i \sim \mathcal{N}(0,\Omega)$ for some individual $i$.
The lognormal distribution results from a transformation of a normal random variable.
If $Y \sim \mathcal{N}(\mu,\Sigma)$, then $X = e^Y$ is lognormally distributed, i.e $X \sim \mathcal{LN}(\mu,\Sigma)$. The probability density of the multivariate lognormal distribution is given by
\begin{equation}\label{eq:mvlogn}
p(x|\mu,\Sigma) = \frac{1}{(2\pi)^{d/2}|\Sigma|^{1/2}} \cdot \frac{1}{\prod_k x_k} \cdot e^{-\frac{1}{2} (\text{log} x - \mu)^T\Sigma^{-1}(\text{log} x - \mu)}\,,
\end{equation}
where $\mu \in \mathbb{R}^d$ and $\Sigma \in \mathbb{R}^{d\times d}$ are the parameters of the associated normal distribution and log denotes the natural logarithm.Thus, the prior for the individual parameters is given by
\begin{align*}
p_\Theta(\theta) &= \mathcal{LN}(\theta|\text{log}(\hatthetaTV),\hatOmega)\\
&= \frac{1}{(2\pi)^{d/2}|\hatOmega|^{1/2}} \cdot \frac{1}{\prod_k \theta_k} \cdot e^{-\frac{1}{2} (\text{log} (\theta) - \text{log}(\hatthetaTV))^T\hatOmega^{-1}(\text{log} (\theta) - \text{log}(\hatthetaTV))}\,. 
\end{align*}

\section{Maximum a-posteriori (MAP) estimation }\label{sec:MAP}
In MAP estimation, one seeks the parameter values that maximise the posterior probability
\begin{equation*}
\hat{\theta}^\MAP_n = \underset{\theta}{\arg \max} \ p(\theta|y_{1:n})\,.
\end{equation*}
It is, however, more convenient and numerically more stable to minimise the negative log-posterior instead
\begin{align*}
\hat{\theta}^\MAP_n &= \underset{\theta}{\arg \min} \ - \log p(\theta|y_{1:n})\\
&= \underset{\theta}{\arg \min} \ - \log p(y_{1:n}|\theta) - \log p(\theta)\,.
\end{align*}

Choosing an additive normal residual error model ($Y_j = h_j +\epsilon_j$ with $ \epsilon_j \sim_\iid \mathcal{N}(0,\sigma^2)$) and a lognormal IIV model for the parameters ($\theta_k = \theta_k^{TV}\cdot e^{\eta_k}$ with $\eta_k \sim_\iid \mathcal{N}(0,w^2_k)$) yields
\begin{align*}
\hat{\theta}^\MAP_n = \underset{\theta}{\arg \min} \ &\frac{n}{2}\log (2\pi) + \frac{n}{2} \log \sigma^2+ \frac{1}{2} \sum_{j=1}^n \frac{(y_j - h_j(\theta))^2}{\sigma^2}\\ 
&+\frac{d}{2}\log (2\pi) + \frac{1}{2} \sum_{k=1}^d \log \omega_k^2+ \sum_{k=1}^d \log \theta_k +\frac{1}{2} \sum_{k=1}^d \frac{(\text{log} (\theta_k) -\text{log}(\hatthetaTV_k))^2}{\omega^2_k} 
\end{align*}
with data $y_{1:n}=(y_1,\dots,y_n)^T$ observed up to time point $t_n$. MAP estimation was performed in Matlab R2017b using the interior-point algorithm in fmincon. The Matlab toolbox AMICI \cite{Frohlich2016} was used for simulation of the system of ordinary differential equations (ODEs) and for computations of sensitivities used in gradient calculations (described below).


Gradient descent algorithms can often be improved by providing the gradient and the hessian of the objective function $J(\theta) = - \log p(\theta|\yn)$. The gradient for this specific problem is given by
\begin{align*}
\frac{\partial J(\theta)}{\partial \theta_l} =& - \sum_{j=1}^{n}\frac{(y_j-h_j(\theta))}{\sigma^2} \cdot \frac{\partial h_j(\theta)}{\partial \theta_l} \\
&+\frac{1}{\theta_l} + \frac{(\text{log}(\theta_l)-\text{log}(\hatthetaTV_l))}{\omega_l^2} \cdot \frac{1}{\theta_l}\,,
\end{align*}
and the hessian for $l \neq m$
\begin{align*}
\frac{\partial^2 J(\theta)}{\partial \theta_l \partial \theta_m} =& - \Big( \sum_{j=1}^{n}\frac{(y_j-h_j(\theta))}{\sigma^2} \cdot \frac{\partial^2 h_j(\theta)}{\partial \theta_l \partial \theta_m} - \frac{1}{\sigma^2} \frac{\partial h_j(\theta)}{\partial \theta_l}\frac{\partial h_j(\theta)}{\partial \theta_m}\Big)
\end{align*}
and
\begin{align*}
\frac{\partial^2 J(\theta)}{\partial \theta_l^2} =& - \Big( \sum_{j=1}^{n}\frac{(y_j-h_j(\theta))}{\sigma^2} \cdot \frac{\partial^2 h_j(\theta)}{\partial \theta_l ^2} - \frac{1}{\sigma^2} \frac{\partial h_j(\theta)}{\partial \theta_l}\frac{\partial h_j(\theta)}{\partial \theta_l}\Big)\\
&+\frac{1}{\theta_l^2} \cdot \Big[ \frac{1}{w_l^2} \Big(1 - \log(\theta_l)+\log(\hatthetaTV_l) \Big)-1 \Big]\,.
\end{align*}


Note that $\frac{\partial h_i(\theta)}{\partial \theta_l} = s^h_l$ are the output sensitivities, which are given by
\begin{equation*}
s^h_l=\frac{\partial h(x,\theta)}{\partial x}s^x_l + \frac{\partial h(x,\theta)}{\partial \theta_l}\,,
\end{equation*}
using the sensitivities of the states
\begin{equation*}
\frac{\partial s^x_l}{\partial t}=\frac{\partial f(x,\theta)}{\partial x}s^x_l + \frac{\partial f(x,\theta)}{\partial \theta_l}\,, \qquad s^x_l(0)=\frac{\partial x_0(\theta)}{\partial \theta_l}\,.
\end{equation*}
For the computation of the state sensitivities the extended system of ODEs needs to be solved
\begin{align*}
\dot{x} &= f(x,\theta), &x(0) &= x_0(\theta)\\
\dot{s}^x_l &= \frac{\partial f(x,\theta)}{\partial x}s^x_l + \frac{\partial f(x,\theta)}{\partial \theta_l}\,, &s^x_l(0)&=\frac{\partial x_0(\theta)}{\partial \theta_l}\,. \\
\end{align*}
Alternatively, the gradient could be computed via adjoint sensitivity analysis which is more efficient for models with large number of states and parameters.

Since the hessian matrix requires the computation of the second-order sensitivities $\frac{\partial^2 h_i(\theta)}{\partial \theta_l \partial \theta_m}$, which is computationally expensive, often the (expected) Fisher information matrix $\FIM$ is used as approximation

\begin{equation*}
\mathcal{I}_{lm}(\theta)= -  \sum_{i=1}^{n}  \frac{1}{\sigma^2}  \cdot \frac{\partial h_i(\theta)}{\partial \theta_l}  \cdot \frac{\partial h_i(\theta)}{\partial \theta_m} \,,
\end{equation*}
and
\begin{equation*}
\mathcal{I}_{ll}(\theta)= - \Bigg( \sum_{i=1}^{n}  \frac{1}{\sigma^2}  \cdot \frac{\partial h_i(\theta)}{\partial \theta_l}  \cdot \frac{\partial h_i(\theta)}{\partial \theta_l} 
+ \frac{1}{\omega^2_l \theta^2_l}\cdot \Big(\text{log}(\hatthetaTV_l)-\text{log}(\theta_l)+1\Big) \Bigg)\,.
\end{equation*}

%
%

\begin{figure}
\centering
\includegraphics[scale=0.6]{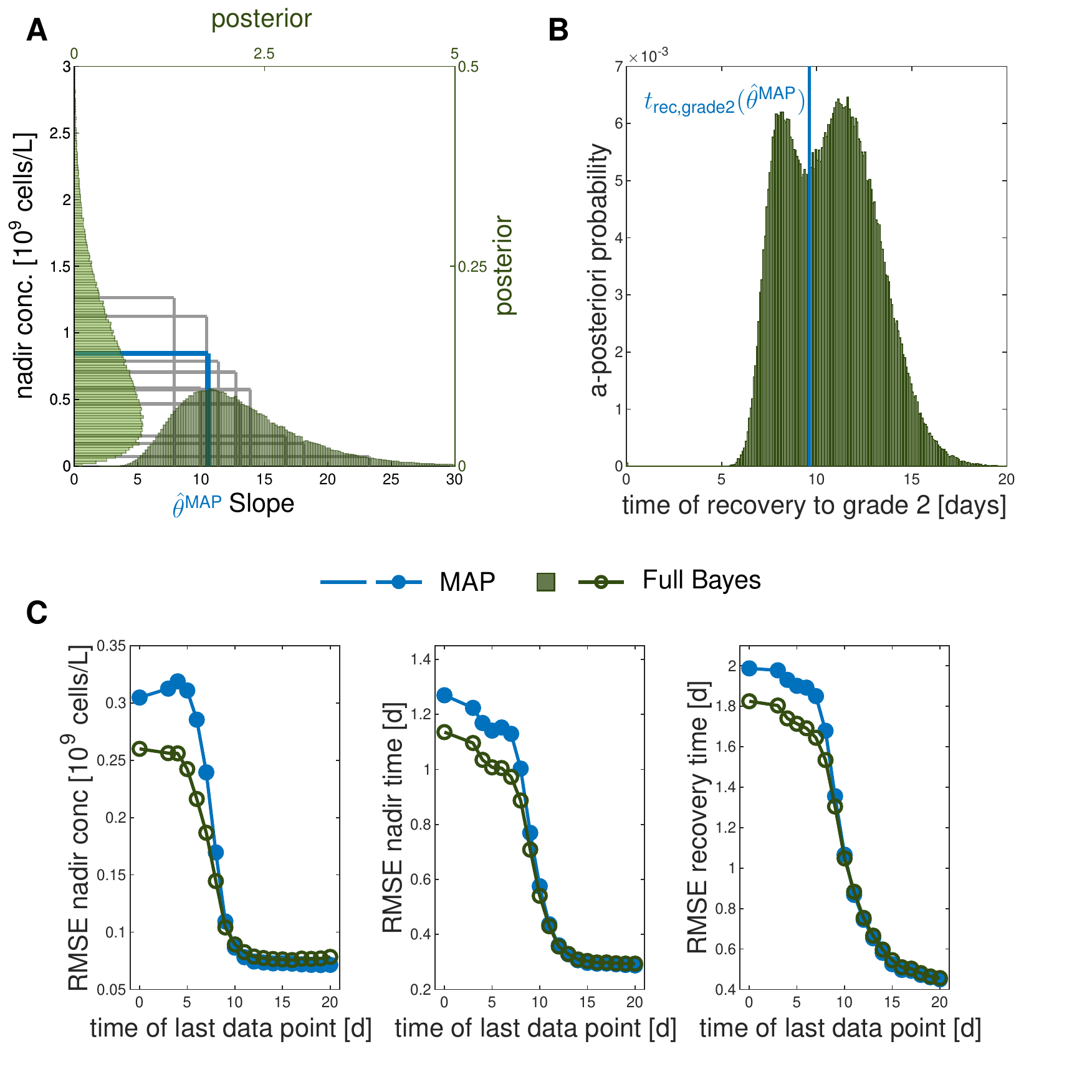}
\caption{ \textbf{Illustration of the unfavourable propoerties of MAP estimation with regard to reliable decision support}. \textbf{A} The MAP estimate (blue line) does not correctly transform under a nonlinear mapping to the most probable nadir concentration (based on \cite[Figure 5.2]{Murphy2012}). The posterior of the parameter '$\Slope$' is depicted for an exemplary patient after four data points $y_{1:4}$ were observed on the x-axis and the corresponding a-posteriori probability of the nadir concentration on the y-axis (same scenario as in Figure~3). \textbf{B} The mode is not preserved under transformation. The same scenario was considered as for part A.  \textbf{C} Root mean squared error (RMSE) of selected statistics. Comparison of the accuracy of the computed statistics $\cnadir,\tnadir$ and $\trec0$ based on MAP estimation and full Bayesian inference (SIR using $S=10^3$ samples). The RMSE was computed across the whole considered virtual population $N=100$.}
\label{fig:MAPvsFBayesABC}
\end{figure}

It was found that the MAP estimate does not correctly transform under a nonlinear mapping to the most probable observation/quantity of interest. As pharmacometric PK/PD models are often nonlinear, this is a major drawback for decision support in MIPD. \SFigref{fig:MAPvsFBayesABC} (A) shows the posterior of the drug effect parameter `Slope' on the x-axis and the a-posteriori probability of the nadir concentration on the y-axis. The blue line shows the MAP estimate for the parameter `Slope' and links to the MAP-predicted nadir concentration, which clearly does not correspond to the mode of the a-posteriori probability distribution of the nadir concentration. In addition it is shown how some randomly chosen samples transform to the nadir concentration. Note, however that the nadir concentration does of course not only depend on the `Slope' but also on the other parameters. \SFigref{fig:MAPvsFBayesABC} (B) demonstrates the same observation for a different quantity of interest, the time of recovery to grade 2, which shows a bimodal posterior distribution.
Further we considered as a statistical measure of accuracy the root mean squared error (RMSE) between the model-predicted outcome $T_i(\Sample_n)$ given data $\yn$ for individual $i$ and the reference  outcome $\Stat^\text{ref}$ (for which the data was simulated)
\begin{equation*}
\text{RMSE}(\Stat)_n = \sqrt{\frac{1}{N} \sum_{i=1}^N (T_i(\Sample_n)- \Stat_i^\text{ref})^2}\,.
    \end{equation*}
  \SFigref{fig:MAPvsFBayesABC} (C) shows the prediction accuracy of the point-estimates of MAP-estimation and of the Full Bayesian approach over time. As more data points are taken into account the RMSE decreases for both estimators. In the beginning of the cycle the Full Bayesian approach shows increased accuracy across all considered quantities of interest.

\section{Normal Approximation (NAP)}\label{sec:NAP}
To quantify the uncertainty associated to a MAP estimate one may consider a quadratic approximation to the log posterior at its mode (following \cite{Gelman2014}). 
A Taylor expansion about the MAP estimate results in
\begin{align*}
\log p(\theta|\yn) &\approx \underbrace{\log p(\MAPest |\yn)}_{=\text{const.}} + (\theta - \MAPest)^T \underbrace{\Big[ \frac{d \log p(\theta|\yn)}{d\theta} \Big]_{\theta = \MAPest }}_{=0}\\
 &+ \frac{1}{2} (\theta - \MAPest)^T \Big[ \frac{d^2 \log p(\theta|\yn)}{d\theta^2} \Big]_{\theta = \MAPest} (\theta - \MAPest )\,.
\end{align*}
If we take the exponential on both sides and normalise, we get as normal approximation to the posterior at the MAP estimate
\begin{equation*}
p(\theta|\yn) \approx \mathcal{N}\Big(\MAPest,\Big[ - \frac{d^2 \log p(\theta|\yn)}{d\theta^2} \Big]^{-1}_{\theta = \MAPest}\Big)\,.
\end{equation*}
The inverse of the variance can be decomposed using Bayes' formula
\begin{equation}\label{eq:decompBayes}
- \frac{d^2 }{d\theta^2} \log p(\theta|\yn) = - \frac{d^2}{d\theta^2} \log p(\yn|\theta) - \frac{d^2}{d\theta^2} \log p(\theta)\,,
\end{equation}
where we retrieve the total observed Fisher information matrix (FIM)
\begin{equation*}
\mathcal{I}^{\mathrm{likelihood}}(\theta)  =  - \frac{d^2}{d\theta^2}  \log p(\yn|\theta) \,.
\end{equation*}
If the definition for the observed FIM of the likelihood is transferred to prior and posterior, the previous decomposition \Eqref{eq:decompBayes} can be written as
\begin{equation*}
\mathcal{I}^{\mathrm{post}}(\theta) = \mathcal{I}^{\mathrm{likelihood}}(\theta) + \mathcal{I}^{\mathrm{prior}}(\theta) \,,
\end{equation*}
which allows to rewrite the normal approximation of the posterior as
\begin{equation}\label{eq:SNAP}
p(\theta|\yn) \approx \mathcal{N}(\MAPest,[\mathcal{I}^{\mathrm{post}}(\MAPest) ]^{-1})\,.
\end{equation}
Note, that we defined $\mathcal{I}\big(\theta \big) := \mathcal{I}^\text{post}\big(\theta \big) $ in the main article.
Since the MAP estimator is asymptotically normally distributed in the limit of large sample sizes ($n \rightarrow \infty$) (see \cite[appendix B]{Gelman2014} for a proof), this approximation can be very precise in the case of highly informative data sets.

\subsection{Simulation-based approach}
To propagate the uncertainties from the parameters to the model predictions, we used a simulation-based approach. To this end,  we sampled from the normal distribution \Eqref{eq:SNAP} and subsequently propagated each sample.
\\

Step-by-step description of the algorithm 
 \begin{enumerate}
\item Generate posterior samples $\thetasample$ from $\mathcal{N}(\MAPest,[\mathcal{I}^{\mathrm{post}}(\MAPest) ]^{-1})$ \,.
 \item Compute for each sample $\ypred(t_j,\thetasample)$ for time points of interest $j=1,\dots,m$
 \item Compute quantiles $(h_{\alpha/2},h_{1-\alpha/2})$ for each time point $t_j$
 \end{enumerate}
 Similarly, we can compute samples of the quantities of interest $T^{(s)}=T(\thetasample)$.

\subsection{Delta Method}\label{sec:DeltaMethod}

As an alternative to the simulation-based approach, we may use the delta method to determine the limiting distribution of a differentiable function of the parameters $g(\theta)$ \cite[section 5.5]{Wasserman2000}. In our case $g(\theta) = h(x(t),\theta)$. Using the Delta method, the uncertainties are propagated from the parameters to the observable via the output sensitivities $S^h(\hat{\theta}^\MAP) = \nabla_\theta h_t(\theta)\big\rvert_{\hat{\theta}^\MAP}$  \cite[section 5.5]{Wasserman2000}:
\begin{equation*}
p(h_t(\hat{\theta}^\MAP)|\yn) \approx \mathcal{N} \Big( h(x(t),\hat{\theta}^\MAP),\, S^h(\hat{\theta}^\MAP)^T \Sigma^\MAP S^h(\hat{\theta}^\MAP) \Big)
\end{equation*}
with $\Sigma^\MAP = [\mathcal{I}(\hat{\theta}^\MAP)]^{-1}$. The credible interval (CI) for the prediction is then given by
\begin{equation*}
\textrm{CI}^\alpha = \ypred(x(t),\MAPest) \pm z_{1-\alpha/2} \sqrt{\sigma^2_\textrm{CI}(t)}\,,
\end{equation*}
where $z_{1-\alpha/2}$ is the quantile of the standard normal distribution and $\sigma^2_\textrm{CI}$ is computed using the output sensitivities 
\begin{equation*}
\sigma^2_\textrm{CI}(t) \approx S^h(t,\MAPest)^T\Sigma^\text{MAP}S^h(t,\MAPest)\,.
\end{equation*}
Alternatively, the quantiles of the Student's t distribution could be used as a more conservative choice, see \cite{Kummel2018}.
For the determination of the prediction interval (PI),  the residual variability is additionally taken into account 
\begin{equation*}
\sigma^2_\textrm{PI}(t) \approx S^h(t,\MAPest)^T\Sigma^\text{MAP}S^h(t,\MAPest) + \sigma_\text{RUV}^2\,.
\end{equation*}
Note that, in the main manuscript $\sigma_\text{RUV}^2 = \sigma^2$.
Consequently, the prediction interval is given by
\begin{equation*}
\textrm{PI}^\alpha = \ypred(t,\MAPest) \pm z_{1-\alpha/2} \sqrt{\sigma^2_\textrm{PI}(t)}\,,
\end{equation*}
Since the delta method involves differentiation of $g$, it is not straightforward to apply the method to any desired quantity of interest, e.g. $g(\theta)=\Stat(\theta)= C_\text{nadir}(\theta)$.

\begin{figure}
\centering
\includegraphics[width =1\linewidth]{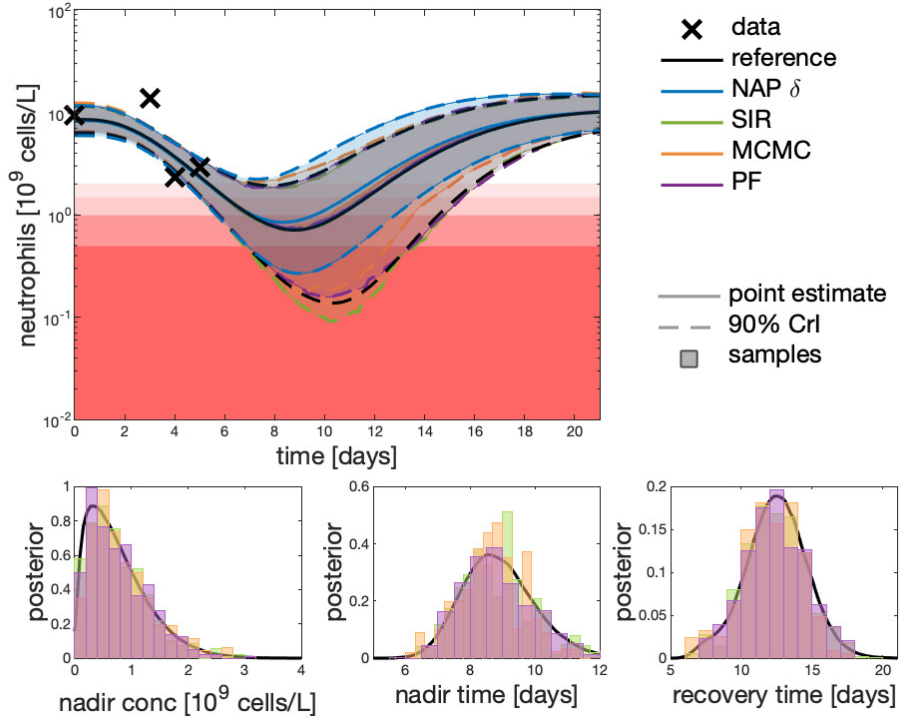}
\caption{Propagating uncertainties using the delta method instead of the simulation based approach.}
\label{fig:NAPDelta}
\end{figure}

The Delta method leads to a similar underestimation of the uncertainty as the simulation-based approach (NAP sim), see \SFigref{fig:NAPDelta}. In addition it is not straightforward to propagate the uncertainty to quantities of interest (therefore not displayed).

\begin{figure}
\centering
\includegraphics[width =1\linewidth]{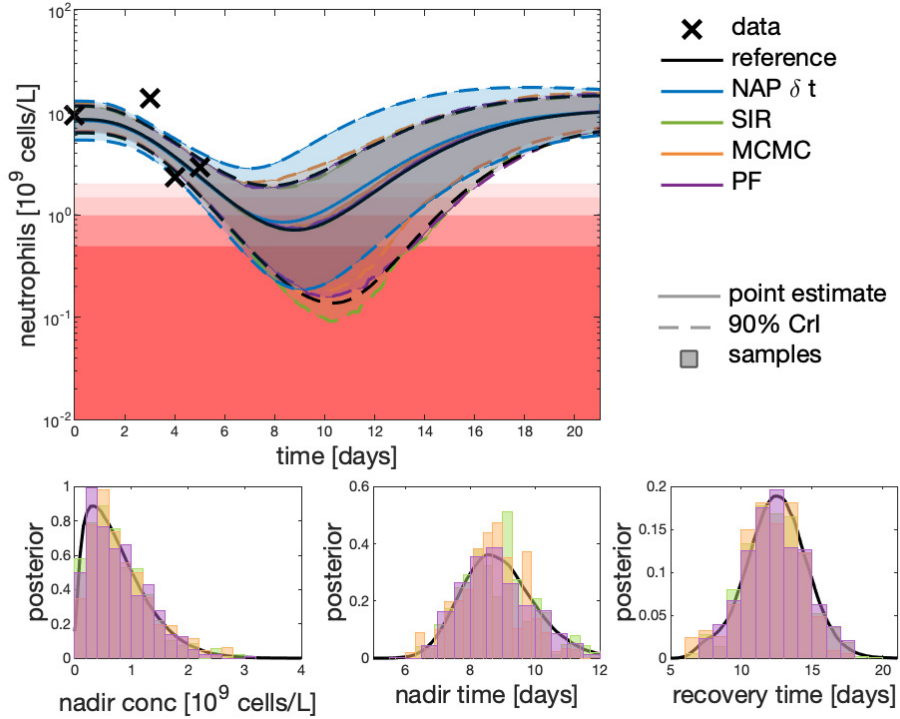}
\caption{Propagating uncertainties with the delta method using Student t quantiles.}
\label{fig:NAPDeltat}
\end{figure}

One option to overcome the underestimation of the uncertainty is the use of quantiles of the Student's t distribution instead of normal quantiles \cite{Kummel2018}. We have used quantiles of the Student's t distribution with $\nu=4$ degrees of freedom (NAP $\delta$ t). The credible intervals show an increased width, but now overestimate the uncertainties regarding subtherapeutic areas (grade 0), see \SFigref{fig:NAPDeltat}. This is also not acceptable as underdosing is in oncology highly undesireable.

\section{Sampling Importance Resampling (SIR)}\label{sec:SIR}
Our goal is to generate samples from the posterior distribution at time point $t_n$ having observed patient-specific data $\yobs_{1:n}$. For this importance sampling is used in the SIR algorithm.
Let $\tilde{p}(\theta|\yobs)$ be the unnormalised posterior
\begin{equation*}
\tilde{p}(\theta|\yobs_{1:n}) = p(\yobs_{1:n}|\theta) \cdot p(\theta).
\end{equation*}
We use the prior $p(\theta) = p(\theta|\thetaTV(\text{cov}),\Omega)$ as importance function.
This gives for a proposal sample $\tilde{\theta}_n^{(s)}$ the unnormalised weights
\begin{equation*}
\tilde{w}^{(s)}_n = \frac{\tilde{p}(\tilde{\theta}_n^{(s)}|\yn)}{p(\tilde{\theta}_n^{(s)})} = p(\yn|\tilde{\theta}_n^{(s)})\,,
\end{equation*}
which need to be normalised for an approximation of the normalised posterior distribution
\begin{equation*}
w^{(s)}_n = \frac{p(\yn|\tilde{\theta}_n^{(s)})}{\sum_{s'} p(\yn|\tilde{\theta}_n^{(s')})} \,.
\end{equation*}
Step-by-step description of the algorithm 
\begin{enumerate}
\item Sample proposals $\theta_n^{(s)}$ from prior  $ p(\theta|\thetaTV,\Omega)$
\item Compute unnormalised importance weights $\tilde{w}_n^{(s)}$
\item Compute normalised importance weights $w_n^{(s)}$
\item Resample according to normalised importance weights $w_n^{(s)}$
\end{enumerate}
Generally this algorithm needs a large number of samples, especially if there is a large disagreement between prior and posterior. This is computationally very expensive, but can be run in parallel up to the normalisation of the weights. The SIR algorithm is also used in the population analysis context to improve the estimation of the parameter uncertainty distributions \cite{Dosne2016}. Thus, the SIR approach might be an option to also take into account uncertainty in the population parameters (hyper parameters).

\subsubsection*{Reference posterior}
Since the true posterior distribution is analytically intractable, we employ as reference solution the SIR algorithm with a large number of samples ($S=10^6$),  as the algorithm is exact for $S \rightarrow \infty$. 
To validate this reference we have compared it in addition to the posterior derived by the MCMC algorithm using also $S=10^6$ samples with a burn-in of $100$ samples, see \SFigref{fig:ReferencePosterior}.

\newpage
\section{Markov chain Monte Carlo (MCMC)}\label{sec:MCMC}
\begin{figure}
\centering
\includegraphics[width =0.5\linewidth]{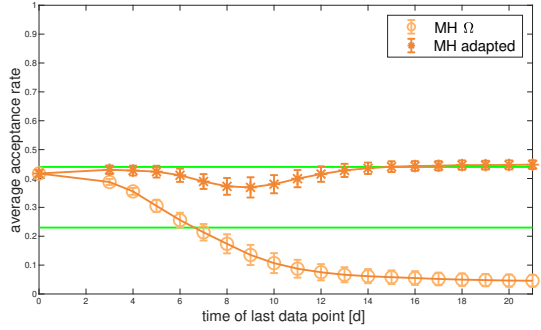}
\caption{Acceptance rate of the Metropolis Hastings algorithm with fixed proposal variance (MH$\Omega$) and with adapted proposal (MH adapted). The green lines mark the area of a good acceptance rate balancing the trade-off between exploring the space and efficiently moving the chain.}
\label{fig:AcceptanceMCMC}
\end{figure}

We used in the presented study an adaptive version of the well-known Metropolis-Hastings algorithm.
In the TDM setting it was previously suggested to use the prior as fixed proposal (independence sampler) \cite{Wakefield1996} .
We have, however, found that as the number of data points increases the rejection rate increases (since the posterior is becoming narrower), leading to inefficient sampling of the posterior. However, positioning the proposal ($\mathcal{LN}(\cdot|\Omega)$) at the just accepted proposal also showed high rejection rates. To counteract low acceptance rates, the proposal was therefore not only moved to the just accepted proposal but was also adapted to the previous posterior sampling distribution, e.g. $\theta_0 =\mathbb{E}[\hat{p}(\log(\theta)|y_{1:n-1}])$ and $\mathcal{LN}(\cdot|\theta_{s-1},\text{Cov}[\hat{p}(\log(\theta)|y_{1:n-1}])$, see \SFigref{fig:AcceptanceMCMC}.

%
Since we employed the Metropolis-Hastings algorithm with a lognormal, thus asymmetric distribution as a proposal, we have to use a correction term in the acceptance probability to account for the asymmetry of the proposal distribution.
Using \Eqref{eq:mvlogn}, we get the following acceptance ratio in the case of a lognormal proposal distribution
\begin{equation} \label{eq:alpha}
\alpha = \frac{p(\theta^*|\yn)}{p(\theta_{s-1}|\yn)} \cdot \frac{\mathcal{LN}(\theta_{s-1}|\theta^*,\hatOmega)}{\mathcal{LN}(\theta^*|\theta_{s-1},\hatOmega)} =  \frac{p(\theta^*|\yn)}{p(\theta_{j-1}|\yn)} \cdot \frac{\prod (\theta^*)_k}{\prod (\theta_{j-1})_k}\,.
\end{equation}
The Markov chain was started at the typical value $\hatthetaTV(\cov)$. Generally, a certain number of samples in the beginning is discarded, a so called burn-in or warm-up. We chose a burn-in of 100 samples in our analysis.

The steps of the Metropolis-Hastings algorithm at time point $t_n$ are
\begin{enumerate}
\item Start chain at $\theta_0 =\hatthetaTV$
\end{enumerate}
For $s=1,\dots,S$
\begin{enumerate}[start=2]
\item Generate proposal $\theta^*$ from the proposal distribution $\mathcal{LN}(\theta_{s-1},\hatOmega)$
\item Accept $\theta^*$ with probability  $\alpha$ defined in \Eqref{eq:alpha}.
\end{enumerate}

One drawback of MCMC approaches is that 
standard MCMC methods cannot be used efficiently in sequential inference context, as for every updated posterior distribution $p(\theta|y_{1:k})$  a new Markov chain has to be generated \cite{DelMoral2006}.

%

%

\section{Particle filter (PF)}\label{sec:PF}

The particle filter belongs to the class of sequential data assimilation methods. In sequential DA methods, the posterior is iteratively updated via Bayes' formula by combining computer-generated Bayesian forecasts with data in real time. The most well-known sequential data assimilation method is the Kalman filter \cite{Kalman1960,KalmanBucy1961}, which relies on the assumptions of linear model dynamics and Gaussian uncertainty, which reduced the problem to only track the mean and variance of the posterior density over time. However, many problems in application
do not satisfy these assumptions. Therefore, particle filters (PF) \cite{Gordon1993} were developed that allow
for non-Gaussian error models and non-linear structural models, which fits the general pharmacometric setting.
In this section some more algorithmic details of the particle filter are described.

\subsection{State augmentation}
Filter algorithms were mainly developed for state estimation with fixed parameters. However, the parameters can be added to an augmented state space $z = (x,\theta)$,
\begin{align*}
\frac{\partial x}{\partial t}(t) &= f(x(t);\theta,u)\\
\frac{\partial \theta}{\partial t}(t) &= 0\,.
\end{align*}
Since we are considering static parameters (within one cycle), the rate of change of parameters is zero.
The filter algorithm was then applied to the augmented state $z(t)=(x(t),\theta(t))$.

\subsection{Resampling strategies}\label{sec:Resampling}
There exist many different strategies on how to resample efficiently and effectively.
The most widely used are multinomial, residual and systematic resampling, see e.g. \cite{Doucet2008}.
For this article,  we applied residual resampling. Resampling can be performed at every step (bootstrap filter) or, more efficiently, only if the effective sample size is smaller than a threshold, e.g. $S_\text{eff}<S/2$.

\subsection{Rejuvenation}
Resampling addresses the problem of weight degeneracy in particle filters, however, it introduces the problem of sample impoverishment. Drawing with replacement among the particles results in many identical particles.
Since the structural model is deterministic with constant parameters, resampled particles will remain identical over time. Particle rejuvenation can be applied to counteract this sample impoverishment \cite{ReichCotter2015}. After each resampling new parameter particles are sampled from a normal distribution centered around the previous parameter values with a small relative variance. Since this only introduces small perturbations in the parameter space, we assumed that $x(\theta) \approx x(\tilde{\theta})$ for a rejuvenation of $\theta$ resulting in $\tilde{\theta}$.

\begin{figure}[t]
\centering
\includegraphics[width = \linewidth]{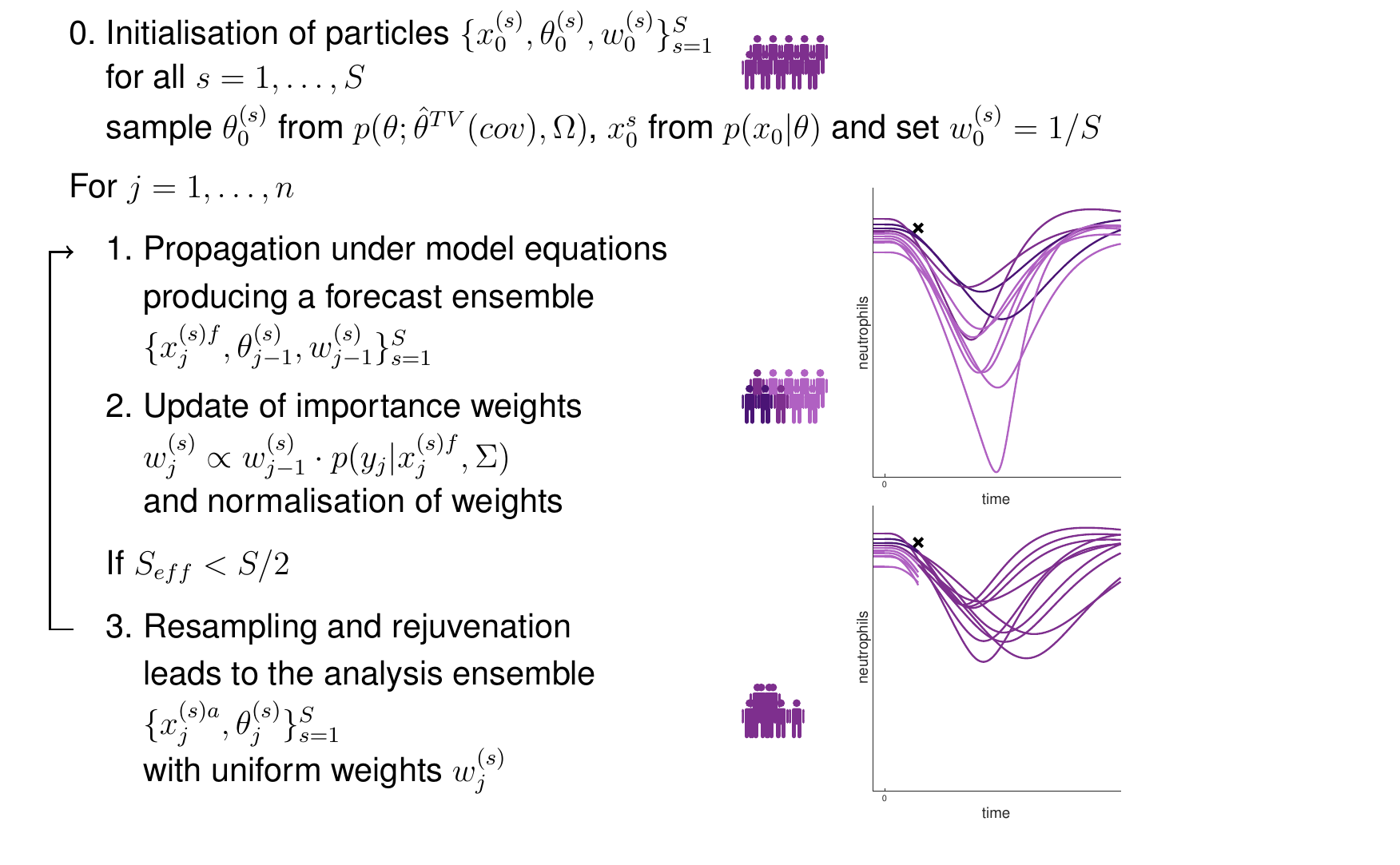}
\caption{\textbf{Step-by-step description of the particle filter}. The different steps of the particle filter are depicted in the context of forecasting the time-course of neutropenia. For illustration purposes the number of virtual individuals is reduced to $S=10$. The different shades represent the importance weights, the darker the colour the higher the importance weight. From the initial virtual subpopulation only five are resampled (resulting in duplicate samples). Rejuvenation introduces new virtual individuals who are similar to the five resampled ones.}
\label{fig:PF_illustration}
\end{figure}

\subsection{Smoothing}
For smoothing over the past prediction the previously predicted paths are resampled along with the states and parameters according to the current particle weights.

\subsection{Alternative sequential DA algorithms}
There exist many extensions, modifications and add-on techniques for particle filters. 
Depending on how the analysis ensemble is generated from the forecast ensemble, different filter algorithms are distinguished, see e.g. \cite{Acevedo2017,Reich2013}.
Among these the class of ensemble transform filters \cite{Reich2013} is very promising as it replaces the stochastic resampling and rejuvenation step of the basic particle filter by a deterministic transformation which allows to ensure certain properties, e.g. 2nd order accuracy \cite{Acevedo2017}. However, in the augmented state space the connection between the parameters and states is lost as larger steps in the parameter steps are undertaken which means that the assumption  $x(\theta) \approx x(\tilde{\theta})$ is no longer valid. 
\newpage

\section{Simulation study: Single cycle Docetaxel}\label{sec:SimStudy1}

In the simulation study we use as prior knowledge a population analysis of a clinical study by Kloft et al. \cite{Kloft2006}. In this section more details about the employed models and parameter estimates is provided. We aimed to be comparable to the setting in \cite{Netterberg2017}.
A virtual population ($N=100$) was generated based on the patient characteristics provided in \cite{Kloft2006}. The covariates AGE and AAG were sampled from a normal distribution with mean given by the median and an estimated variance from the given observed range. The parameter estimates used for the pharmacokinetic (PK) and pharmacodynamic (PD) model are given in \STabref{tab:KloftParameterEstimates}. For the MAP estimation we needed to provide parameter bounds to the optimiser. The lower bounds were taken from the code provided by Netterberg \cite{NetterbergDDMORE} and the upper bounds were tested, so that the optimiser did not reach the bound.

\begin{figure}
\centering
\includegraphics[width =1\linewidth]{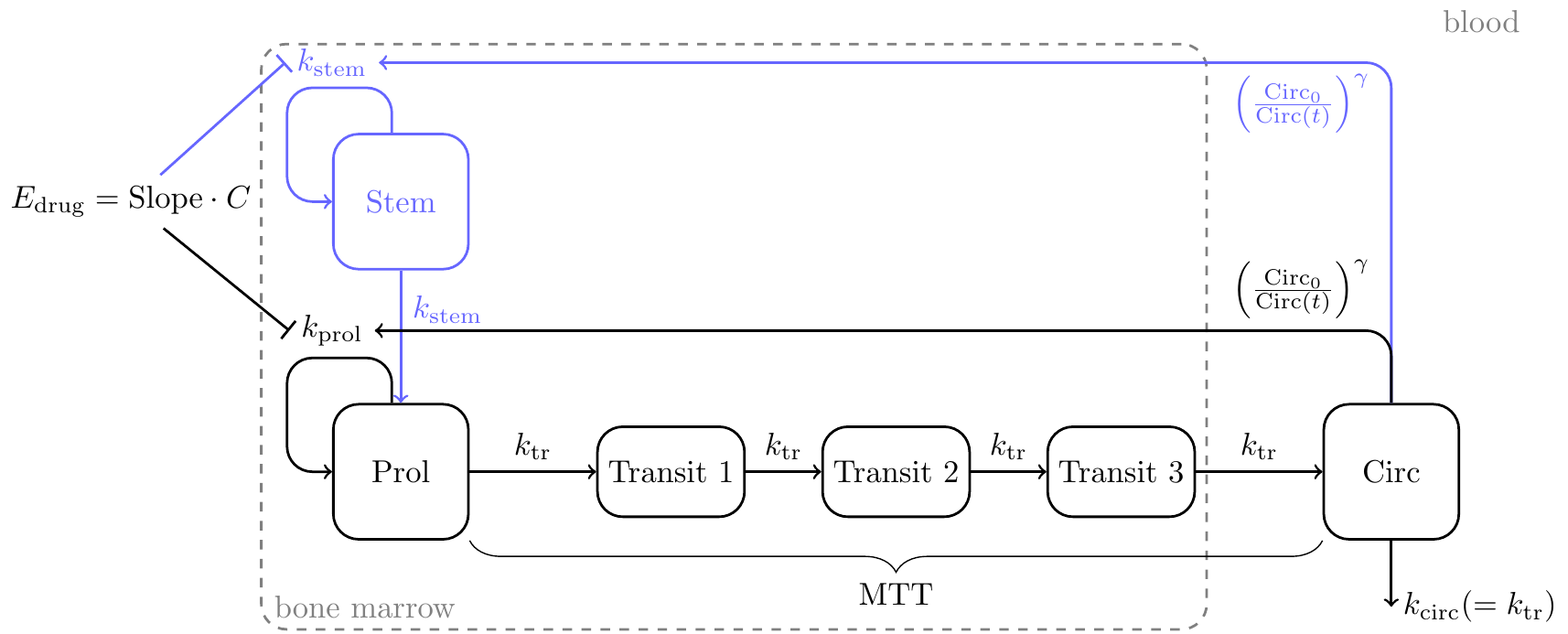}
\caption{\textbf{Structural models for neutropenia}. The gold-standard model for neutropenia, developed by Friberg~et~al. \cite{Friberg2002} (black). Extended bone marrow exhaustion model, which describes cumulative neutropenia over multiple cycles, developed by Henrich~et~al. \cite{Henrich2017} (black and blue). The state variables and parameters of the models are described in the text.}
\label{fig:Structural_Model}
\end{figure}

As PK model for Docetaxel a three compartment model with first-order elimination was employed \cite{Bruno1996}. AAG ($\alpha$1-acid glycoprotein), AGE, BSA (body surface area) and ALB (albumin) were found as covariates on clearance. 
Patient specific parameter values were determined based on covariates. The individual clearance is computed via
\begin{equation*}
\text{CL}_i = \text{BSA}_i \cdot (\mathrm{CL}^\text{TV} + \theta_{\mathrm{CL}\text{-}\mathrm{AAG}} \cdot \text{AAG}_i + \theta_{\mathrm{CL}\text{-}\mathrm{AGE}} \cdot \text{AGE}_i + \theta_{\mathrm{CL}\text{-}\mathrm{ALB}} \cdot \text{ALB}^\text{TV}) \cdot (1- \theta_{\mathrm{CL}\text{-}\mathrm{HEP12}} \cdot \text{HEP12})  \,,
\end{equation*}
where we used $\text{ALB}^\text{TV} = 41\,g/\L$ and set HEP12=0 (i.e. no elevated hepatic enzymes). The parameter estimates were taken from \cite{Bruno1996}, see Supplement \Tabref{tab:BrunoParameterEstimates} and the system of ODEs for the PK model is given by

\begin{align*}
\frac{d \Cent}{dt} &= u(t) - k_{10}\Cent + k_{21} \text{Per1} - k_{12} \text{Cent} + k_{31}\text{Per2} - k_{13}\Cent\,, &\Cent(0) = 0\hphantom{\,.} \\
\frac{d \text{Per1}}{dt} &= k_{12}\Cent - k_{21}\text{Per1}\,, &\text{Per1}(0)=0\hphantom{\,.}\\
\frac{d \text{Per2}}{dt} &= k_{13}\Cent - k_{31}\text{Per2}\,, &\text{Per2}(0)=0\,.
\end{align*}
%
\begin{table}[h]
\begin{center}
\begin{tabular}{lll}
\hline
\multicolumn{3}{c}{Structural submodel}\\
\hline
$\mathrm{V}$ 	& 	8.31 				& $[\L]$\\
$\mathrm{CL}^\text{TV}$ & 	22.1 				& $[\L/h]$\\
$k_{10}$		&	$\mathrm{CL}/\mathrm{V}$ 	& $[1/h]$\\
$k_{12}$		&	1.07 				& $[1/h]$ \\
$k_{21}$		&	1.74 				& $[1/h]$ \\
$k_{13}$		&	1.28				& $[1/h]$ \\
$k_{31}$     	& 	0.0787 			& $[1/h]$ \\
\hline
\multicolumn{3}{c}{Covariate submodel}\\
\hline
$\theta_{\mathrm{CL}\text{-}\mathrm{AAG}}$ &  -3.55 &\\
$\theta_{\mathrm{CL}\text{-}\mathrm{AGE}}$ &  -0.095 &\\
$\theta_{\mathrm{CL}\text{-}\mathrm{ALB}}$ &  \hphantom{-}0.225 &\\
\end{tabular}
\end{center}
\caption{Pharmacokinetic parameter estimates for Docetaxel \cite{Bruno1996}.}
\label{tab:BrunoParameterEstimates}
\end{table}

The semi-mechanistic model by Friberg et al. \cite{Friberg2002} describes chemotherapy-induced neutropenia and consists of five compartments. The proliferating compartment ($\Prol$) represents rapidly dividing progenitor cells in the bone marrow, which replicate with rate $\kprol$.
Three transit compartments ($\Transit~1\text{-}3$) approximate the maturation chain of progenitor cells in the bone marrow to differentiated neutrophils in the systemic circulation, with transition rate $\ktr$.
Circulating neutrophils in blood ($\Circ$) are the part of the system that can be observed. Thus, the state vector for the model is given by 
\begin{equation*}
 x(t) = \Big(\Prol(t), \Transit1(t), \Transit2(t),\Transit3(t),\Circ(t)\big)^T\,,
\end{equation*}
with observable
\begin{equation*}
 h(t) = \Circ(t)\,,
\end{equation*}
and initial conditions
\begin{equation*}
\Prol(t_0)= \Transit~1(t_0)=\Transit~2(t_0)=\Transit~3(t_0)=\Circ(t_0)=\Circ_0\,.
\end{equation*}
The cytotoxic effect of anticancer drugs is implemented linearly on the proliferation rate of susceptible progenitor cells, so the effect is proportional to the drug concentration
\begin{equation*}
\Edrug(t) = \Slope \cdot \Cdrug(t)\,,
\end{equation*}
which is the input to the system
\begin{equation*}
u(t) = \Cdrug(t)\,,
\end{equation*}
where the time-evolution is given by the corresponding PK model for Docetaxel. The effect of the drug on the proliferation rate of the Prol compartement is described by a feedback, which leads to an increase in production of white blood cells if the number of circulating neutrophils is decreasing, and vice versa
\begin{equation*}
\mathrm{Feedback}(t) = \bigg(\frac{\Circ_0}{\Circ (t)}\bigg)^\gamma \,.
\end{equation*}


System parameters include the baseline neutrophil concentration ($\Circ_0$) before the start of treatment, the mean transition time ($\MTT$), which represents the average time for a progenitor cell in the bone marrow to mature to a circulating neutrophil, and the slope parameter ($\mathrm{Slope}$) of the linear inhibitory model for the drug effect. Furthermore, $\gamma$ is the exponent parameter for the feedback model.
Thus, the parameter vector for the model is given by
\begin{equation*}
 \theta = (\Circ_0, \MTT, \Slope,\gamma)^T\,.
 \end{equation*}
 As in \cite{Netterberg2017} we assume $\gamma$ to be fixed. The system of ordinary differential equations (ODEs) reads
 \begin{align*}
\frac{d \Prol}{dt} &= k_{\text{prol}}\Prol \cdot (1-\Edrug) \cdot \left( \frac{\Circ_0}{\Circ} \right)^\gamma - k_\text{tr} \Prol\,, &\Prol(0) = \Circ_0\hphantom{\,.} \\
\frac{d \text{Transit1}}{dt} &= k_\text{tr}\Prol - k_\text{tr} \text{Transit1}\,, &\text{Transit1}(0)=\Circ_0\hphantom{\,.}\\
\frac{d \text{Transit2}}{dt} &= k_\text{tr}\text{Transit1} - \ktr \text{Transit2}\,, &\text{Transit2}(0)=\Circ_0\hphantom{\,.}\\
\frac{d \text{Transit3}}{dt} &= k_\text{tr}\text{Transit2} - \ktr \text{Transit3}\,, &\text{Transit3}(0)=\Circ_0\hphantom{\,.}\\
\frac{d \Circ}{dt} &= \ktr \text{Transit3} - k_\text{circ} \Circ \,, &\text{Circ}(0)=\Circ_0\,.
\end{align*}



\begin{table}
\centering
\begin{tabular}{lll}
\hline
\multicolumn{3}{c}{Structural submodel}\\
\hline
$\Circ_0$	&	5.22 & $[10^9 \cells/\L]$\\
$\MTT	$	&	84.2 & $[h]$ \\
$\Slope$	&	15.6 & $[\L/\mu \text{mol}]$\\
$\gamma$		&	0.145 & $[ ]$ \\
\hline
\multicolumn{3}{c}{Covariate submodel}\\
\hline
$\theta_{\Circ_0\text{-}\mathrm{AAG} \leq 1.34}$ &  \hphantom{-}0.175 & \\
$\theta_{\Circ_0\text{-}\mathrm{AAG} > 1.34}$ &  \hphantom{-}0.495 & \\
$\theta_{\Circ_0\text{-}\mathrm{SEX}}$ &  -0.121 & \\
$\theta_{\Circ_0\text{-}\mathrm{PERF}}$ &  \hphantom{-}0.131 & \\
$\theta_{\Circ_0\text{-}\mathrm{PC}}$ &  -0.147 & \\
$\theta_{\Slope \text{-}\mathrm{AAG}}$ &  -0.351 & \\
\hline
\multicolumn{2}{c}{Statistical submodel}\\
\hline
$\sigma^2$		& 0.180 & \\
$\omega^2_{\Circ_0}$ & 0.0606 &\\
$\omega^2_{\MTT}$ & 0.0194 &\\
$\omega^2_{\Slope}$ & 0.122 &\\
$\omega^2_{\gamma}$ & 0.0223 &\\
\end{tabular}
\caption{Parameter estimates for the Friberg-model and Docetaxel taken from \cite{Kloft2006}}
\label{tab:KloftParameterEstimates}
\end{table}

\begin{table}
\centering
\begin{tabular}{lll}
\hline
parameter & lower bound & upper bound\\
\hline
$\Circ_0$ $[10^9 \cells/\L]$	& 2 & 30\\
$\MTT	$	$[h]$  &	10 & 250\\
$\Slope$	$[\L/\mu \text{mol}]$ &	0.01 & 60\\
\end{tabular}
\caption{Bounds for MAP estimation. Lower bounds were taken from \cite{NetterbergDDMORE}}
\label{tab:BoundsMAP}
\end{table}

\begin{figure}
\centering
\includegraphics[scale=0.6]{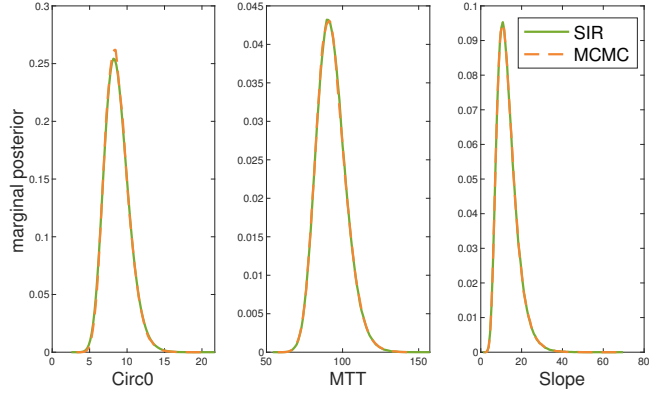}
\caption{Comparison of reference posterior. The reference posterior was derived by the SIR algorithm and with the MCMC algorithm using $S=10^6$ samples. }
\label{fig:ReferencePosterior}
\end{figure}

\newpage
\section{Simulation study: Multiple cycles Paclitaxel}\label{sec:SimStudy2}
The multiple cycle simulation study is based on a population analysis \cite{Henrich2017} of the CEPAC-TDM study data. For the data simulation it was assumed that every third day the neutrophil counts were assessed for a patient undergoing chemotherapy. The data was simulated based on a model prediction including interoccasion variability.
Paclitaxel pharmacokinetics were described by a three compartment model with nonlinear distribution to the first peripheral compartment and nonlinear elimination \cite{Joerger2012}. The system of ODEs reads
\begin{align*}
\frac{d \Cent}{dt} &= u(t) - \frac{\mathrm{VM}_\mathrm{EL} \cdot C_1}{\mathrm{KM}_\mathrm{EL}+C_1} + k_{21} \text{Per1} - \frac{ \mathrm{VM}_\mathrm{TR} \cdot C_1}{\mathrm{KM}_\mathrm{TR}+C_1}  + k_{31}\text{Per2} - k_{13}\Cent\,, &\Cent(0) = 0 \\
\frac{d \text{Per1}}{dt} &= \frac{\mathrm{VM}_\mathrm{TR} \cdot C_1}{\mathrm{KM}_\mathrm{TR}+C_1} - k_{21}\text{Per1}\,, &\text{Per1}(0)=0\\
\frac{d \text{Per2}}{dt} &= k_{13}\Cent - k_{31}\text{Per2}\,, &\text{Per2}(0)=0\\
\end{align*}
where $C_1(t) = \Cent(t)/V_1$.
We used the parameter values by \cite{Henrich2017}, see \STabref{tab:PaclitaxelParameterEstimates}. 
The following covariate model for $\mathrm{VM}_\mathrm{EL}$ was given as
\begin{equation*}
\mathrm{VM}_\mathrm{EL,TV,i} = \mathrm{VM}_\mathrm{EL,pop} \cdot \Big(\frac{\mathrm{BSA}_i}{1.8 m^2}\Big)^{\theta_{\mathrm{VM}_\mathrm{EL}\text{-}\mathrm{BSA}}} \cdot \Big(\theta_{\mathrm{VM}_\mathrm{EL}\text{-}\mathrm{SEX}}\Big)^\mathrm{SEX_i} \cdot \Big(\frac{\mathrm{AGE}_i}{56 \mathrm{years}}\Big)^{\theta_{\mathrm{VM}_\mathrm{EL}\text{-}\mathrm{AGE}}} \cdot \Big(\frac{\mathrm{BILI}_i}{7 \mu \mathrm{mol}/\L}\Big)^{\theta_{\mathrm{VM}_\mathrm{EL}\text{-}\mathrm{BILI}}}\,
\end{equation*}
relating patient characteristics, BSA (body surface area), SEX, AGE and BILI (total bilirubin concentration) to the maximum elimination capacity ($\mathrm{VM}_\mathrm{EL}$). 
In addition to inter-individual variability and residual variability, interoccasion variability was included on the two parameters $V_1$ and $\mathrm{VM}_\mathrm{EL}$, where each start of a new cycle is defined as an occasion,
\begin{equation*}
\theta_{i,o} = \theta_\mathrm{TV,i} \cdot e^{\eta_i + \kappa_{i,o}}\,, \qquad \kappa_{i,o} \iidsim \mathcal{N}(0,\Pi^2)\,.
\end{equation*}



\begin{table}
\begin{center}
\begin{tabular}{lll}
\hline
\multicolumn{3}{c}{Structural submodel}\\
\hline
$\mathrm{V_1}$ 				& 	10.8 		& $[\L]$\\
$\mathrm{V_3}$ 				& 	301 		& $[\L]$\\
$\mathrm{KM}_\mathrm{EL}$ 		& 	0.667 	& $[\mu \M]$\\
$\mathrm{VM}_\mathrm{EL,pop}$ 	& 	35.9  	& $[\mu \mathrm{mol}/h]$\\
$\mathrm{KM}_\mathrm{TR}$ 		& 	1.44	 	& $[\mu \M]$\\
$\mathrm{VM}_\mathrm{TR}$ 		& 	175   	& $[\mu \mathrm{mol}/h]$\\
$k_{21}$						&	1.12 	 	& $[1/h]$\\
$Q$							&	16.8		& $[1/h]$ \\
\hline
\multicolumn{3}{c}{Covariate submodel}\\
\hline
$\theta_{\mathrm{VM}_\mathrm{EL}\text{-}\mathrm{BSA}}$ &  1.14 &\\
$\theta_{\mathrm{VM}_\mathrm{EL}\text{-}\mathrm{SEX}}$ &  1.07 &\\
$\theta_{\mathrm{VM}_\mathrm{EL}\text{-}\mathrm{AGE}}$ &  -0.447 &\\
$\theta_{\mathrm{VM}_\mathrm{EL}\text{-}\mathrm{BILI}}$  &  -0.0942 &\\
\hline
\multicolumn{3}{c}{Statistical submodel IIV}\\
\hline
$\omega^2_{V_3}$ 					&  0.1639 &\\
$\omega^2_{\mathrm{VM}_\mathrm{EL}}$ &  0.0253 &\\
$\omega^2_{\mathrm{KM}_\mathrm{TR}}$ &  0.3885 &\\
$\omega^2_{\mathrm{VM}_\mathrm{TR}}$ &  0.077 &\\
$\omega^2_{k_{21}}$ 					&   0.008  &\\
$\omega^2_{Q}$ 					&  0.1660 &\\
\hline
\multicolumn{3}{c}{Statistical submodel IOV}\\
\hline
$\pi^2_{V_1}$ 					&   0.1391 &\\
$\pi^2_{\mathrm{VM}_\mathrm{EL}}$ 					&  0.0231&\\
\hline
\multicolumn{3}{c}{Statistical submodel RV}\\
\hline
$\sigma^2$ 						& 0.0317& \\
\end{tabular}
\end{center}
\caption{Pharmacokinetic parameter estimates for the anticancer drug Paclitaxel \cite{Henrich2017}.}
\label{tab:PaclitaxelParameterEstimates}
\end{table}

Neutropenia has been observed to worsen over several treatment cycles, i.e. the lowest neutrophil concentration (nadir concentration) and the maximum neutrophil concentration decreases over multiple cycles. Bone marrow exhaustion could be one explanation of this cumulative behaviour, which means that the long-term recovery of the bone marrow is also affected.
The standard model for neutropenia \cite{Friberg2002} does not take such a long-term effect into account.
Henrich et al. \cite{Henrich2017} therefore expanded the model by dividing the proliferating compartment into another stem cell compartment, describing pluripotent stem cells with reduced proliferation rate. This model extension made it possible to capture the cumulative long-term effect.

The proliferation rate constants for the proliferating compartement $\Prol$ and for the stem cell compartement $\Stem$ are given by
\begin{align*}
\kprol &= \ftr \cdot \ktr \\
k_\mathrm{stem} &= (1-\ftr) \cdot \ktr \,, \\
\end{align*}
respectively, where $\ftr$ is the fraction of input in the $\Prol$ compartment via proliferation within the compartment.
As the drug effect is proportional to the proliferation rate constant, the stem cells are less affected by the treatment than the progenitor cells.

 \begin{align*}
 \frac{d \Stem}{dt} &= k_{\text{stem}}\Stem \cdot (1-\Edrug) \cdot \Big( \frac{\Circ_0}{\Circ} \Big)^\gamma - k_\text{trStem} \Stem \,, &\Stem0) = \Circ_0 \\
\frac{d \Prol}{dt} &= k_{\text{prol}}\Prol \cdot (1-\Edrug) \cdot \Big( \frac{\Circ_0}{\Circ} \Big)^\gamma + k_\text{trStem} \Stem - k_\text{tr} \Prol\,, &\Prol(0) = \Circ_0\\
\frac{d \text{Transit1}}{dt} &= k_\text{tr}\Prol - k_\text{tr} \text{Transit1}\,, &\text{Transit1}(0)=\Circ_0\\
\frac{d \text{Transit2}}{dt} &= k_\text{tr}\text{Transit1} - \ktr \text{Transit2}\,, &\text{Transit2}(0)=\Circ_0\\
\frac{d \text{Transit3}}{dt} &= k_\text{tr}\text{Transit2} - \ktr \text{Transit3}\,, &\text{Transit3}(0)=\Circ_0\\
\frac{d \Circ}{dt} &= \ktr \text{Transit3} - k_\text{circ} \Circ \,, &\text{Circ}(0)=\Circ_0\\
\end{align*}

The baseline parameter $\Circ_0$ was inferred from the baseline data point ( baseline method 2 \cite{Dansirikul2008})
\begin{equation*}
\Circ_{0,i}= y_{\Circ_0} \cdot e^{\theta_\textrm{RV} \cdot \eta_{\Circ_0,i}}\,, \qquad \eta_{\Circ_0,i} \sim \mathcal{N}(0,1)\,.
\end{equation*}

\begin{table}[bt]
\begin{center}
\begin{tabular}{lll}
\hline
\multicolumn{3}{c}{Structural submodel}\\
\hline
$\Circ_0$	&	baseline method \cite{Dansirikul2008} & $[10^9 \cells/\L]$\\
$\MTT$	&	145 & $[h]$ \\
$\Slope$	&	13.1 & $[\L/\mu \text{mol}]$\\
$\gamma$		&	0.257 & $[ ]$\\
$\mathrm{ftr}$     	& 	0.787 &$[ ]$ \\
\hline
\multicolumn{3}{c}{Statistical submodel}\\
\hline
$\sigma^2$		& 0.2652 &\\
$\omega^2_{\Slope}$ & 0.2007& \\
\end{tabular}
\end{center}
\caption{Parameter estimates for the bone marrow exhaustion model and the anticancer drug Paclitaxel .}
\label{tab:HenrichParameterEstimates}
\end{table}

In the bone marrow exhaustion model we have in addition to inter-individual variability (IIV) inter-occasion variability (IOV).
The parameter vector, thus, contains parameter values, which are constant across occasions (cycles) $\thetaIIV$ and parameters that are specific for each cycle $c$, $\thetaIOV_c$.
\begin{equation}
\theta_c = \underbrace{e^{\log(\hatthetaTV) + \eta}}_{=\thetaIIV} \cdot \underbrace{e^{\kappa_c}}_{=\thetaIOV_c}\,.
\end{equation}
The $\thetaIIV$ parameters need to be learned across all cycles and the cycle specific parameters $\thetaIOV_c$ based on the data observed in cycle $c$, $y_{1:n_c}=(y_1,\dots,y_{n_c})^T$. The size of the parameter vector that needs to be estimated will, therefore, grow with every occasion if the whole data is processed in a batch. Assuming independence between the inter-individual and interoccasion variability, the objective function for the MAP estimation is then given by
\begin{align*}\label{eq:MAP_objFctIOV}
\hat{\theta}^\MAP_n = \underset{\thetaIIV, \thetaIOV}{\arg \min} \  &\frac{1}{2}\Big(  \sum_{c=1}^{C} \sum_{j=1}^{n_c}\frac{\big(y_{j}-h_j(\theta) \big)^2}{\sigma^2}\\
 &+ 2 \sum_{k=1}^{n_\thetaIIV} \text{log} (\thetaIIV_k) + \sum_{k=1}^{n_\thetaIIV} \frac{\big(\mathrm{log}(\thetaIIV_k)-\mathrm{log}(\hatthetaTV_k(\textrm{cov}))\big)^2}{\omega_k^2}\\
  &+ 2 \sum_{c=1}^C \sum_{k=1}^{n_\thetaIOV} \text{log} (\thetaIOV_{k,c}) + \sum_{c=1}^C \sum_{k=1}^{n_\thetaIOV} \frac{\big(\mathrm{log}(\thetaIOV_{k,c})-\mathrm{log}(\hatthetaTV_k(\textrm{cov}))\big)^2}{\pi_k^2}\Big)
\end{align*}
for the IIV and IOV model $\theta_{k,c} = \theta_k^{TV}\cdot e^{\eta_k+\kappa_{k,c}}\,,$ with $\eta_k \sim \mathcal{N}(0,w^2_k)$ and $\kappa_{k,c} \sim \mathcal{N}(0,\pi^2_{k})$, for an additive normal residual error model $y_j = h_j +\epsilon_j \,,$ with $ \epsilon_j \sim \mathcal{N}(0,\sigma^2)$ and for data observed up to time point $t_{n}$, i.e. $n = \sum_c n_c$, with $n_c$ the number of observations made in cycle $c$.

The gradient with respect to the IIV parameters is then given by
\begin{align*}
\frac{\partial J(\theta)}{\partial \thetaIIV_l} =& - \sum_{j=1}^{n}\frac{(y_j-h_j(\theta))}{\sigma^2} \cdot \frac{\partial h_j(\theta)}{\partial \thetaIIV_l} \\
&+\frac{1}{\thetaIIV_l} + \frac{(\text{log}(\thetaIIV_l)-\text{log}(\theta^\text{TV}_l))}{\omega_l^2} \cdot \frac{1}{\thetaIIV_l}\,,
\end{align*}
and with respect to the IOV parameters by
\begin{align*}
\frac{\partial J(\theta)}{\partial \thetaIOV_{l,c}} =& - \sum_{j=1}^{n}\frac{(y_j-h_j(\theta))}{\sigma^2} \cdot \frac{\partial h_j(\theta)}{\partial \thetaIOV_l} \\
&+\frac{1}{\thetaIOV_{l,c}} + \frac{(\text{log}(\thetaIOV_l)-\text{log}(\theta^\text{TV}_l))}{\omega_l^2} \cdot \frac{1}{\thetaIOV_{l,c}}\,,
\end{align*}

\begin{figure}

\includegraphics[width =1\linewidth]{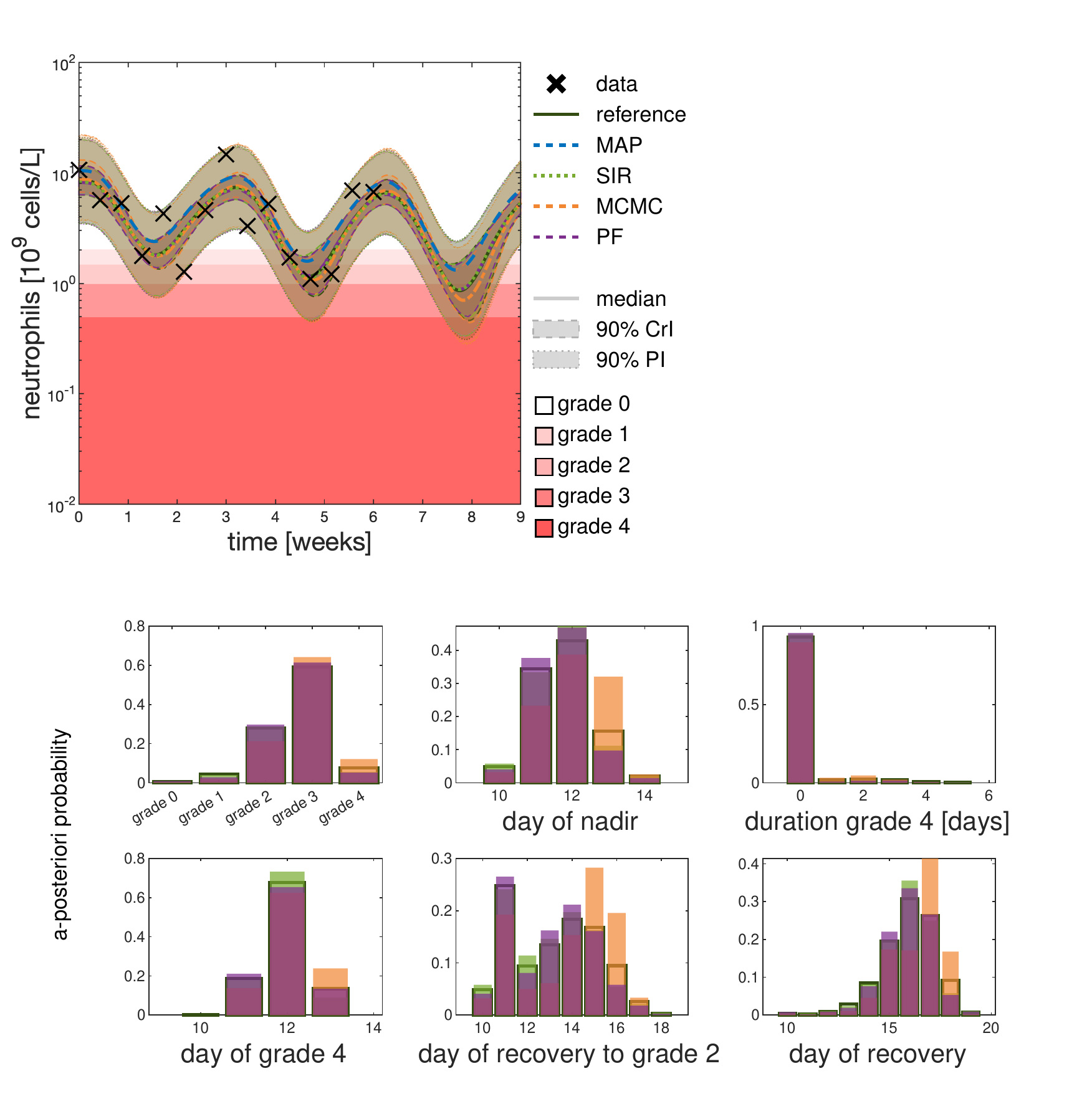}
\caption{\textbf{Comparison of all methods for the standard dose.} The third cycle is forecasted in case the standard dose is given. }
\label{fig:ForecastingNextCycle}
\end{figure}

\SFigref{fig:ForecastingNextCycle} shows a comparison of the different methods in forecasting the third cycle. The scenario corresponds to the situation presented in Figure 4 in the main manuscript. All full Bayesian methods provide almost overlapping credible intervals as well as point estimates (median). However, the MAP-based forecasted trajectory deviates significantly from the point estimates (median) of the full Bayesian methods. 

\begin{figure}

\includegraphics[width =1\linewidth]{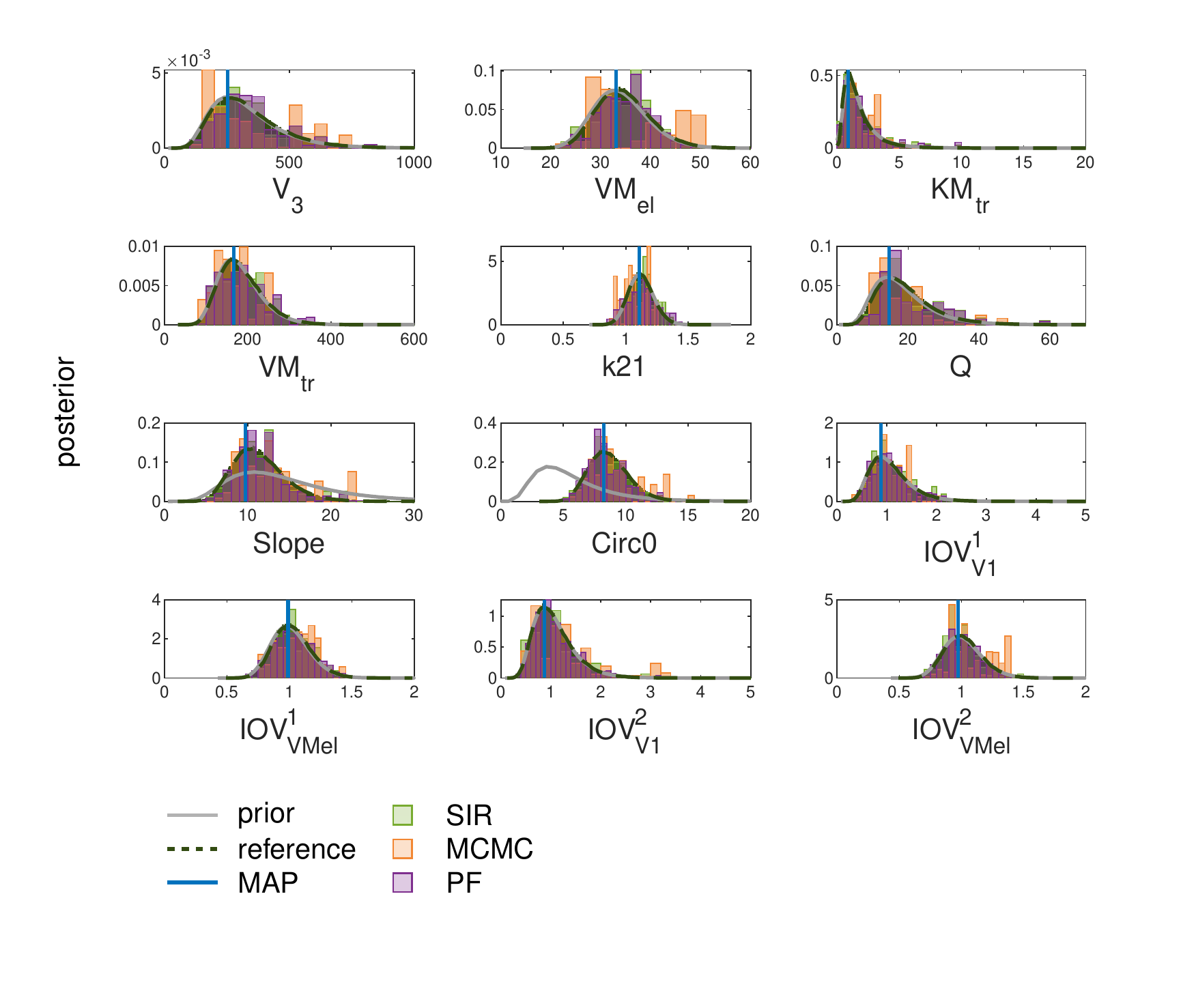}
\caption{\textbf{Posterior inference of the different methods.} Posterior of the parameters for the scenario considered in \Figref{fig:ForecastingNextCycle} }
\label{fig:PosteriorForecastingNextCycle}
\end{figure}

In \SFigref{fig:PosteriorForecastingNextCycle} the posterior approximations are compared to the reference on the level of the parameters. In addition, we can observe the deviation of the posterior from the prior for parameters 'Slope' and 'Circ0'. As we do not consider PK samples the knowledge gain about the PK parameters is limited. All approximations show good agreement with the reference and the MAP estimate is located at the mode of the posterior on the level of the parameters.

\newpage
\bibliographystyle{refstyle}
\bibliography{Maier_etal_2019_literature}